\documentclass{article}

\PassOptionsToPackage{numbers, compress}{natbib}

     \usepackage[preprint]{neurips_2021}

\usepackage[utf8]{inputenc} 
\usepackage[T1]{fontenc}    
\usepackage{hyperref}       
\usepackage{url}            
\usepackage{booktabs}       
\usepackage{amsfonts}       
\usepackage{nicefrac}       
\usepackage{microtype}      
\usepackage{xcolor}         
\usepackage{graphicx}
\usepackage{amsmath}
\usepackage{kotex}
\usepackage{caption}
\usepackage{subcaption}
\usepackage{multirow}
\usepackage[ruled,vlined]{algorithm2e}

\newcommand{\comment}[1]{}

\title{UniTTS: Residual Learning of Unified Embedding Space for Speech Style Control}

\author{
  Minsu Kang, Sungjae Kim, and Injung Kim \\ \\
  Department of Computer Science and Electronic Engineering\\
  Handong Global University\\ 
  \{mskang, 21400110, ijkim\}@handong.edu

}

\begin{document}

\maketitle

\begin{abstract} \label{abstract}
  We propose a novel high-fidelity expressive speech synthesis model, UniTTS, that learns and controls overlapping style attributes avoiding interference. UniTTS represents multiple style attributes in a single unified embedding space by the residuals between the phoneme embeddings before and after applying the attributes. The proposed method is especially effective in controlling multiple attributes that are difficult to separate cleanly, such as speaker ID and emotion, because it minimizes redundancy when adding variance in speaker ID and emotion, and additionally, predicts duration, pitch, and energy based on the speaker ID and emotion. In experiments, the visualization results exhibit that the proposed methods learned multiple attributes harmoniously in a manner that can be easily separated again. As well, UniTTS synthesized high-fidelity speech signals controlling multiple style attributes. The synthesized speech samples are presented at \url{https://anonymous-authors2022.github.io/paper_works/UniTTS/demos/}.

\end{abstract}

\section{Introduction}  \label{introduction}

In recent years, speech synthesis technology has rapidly advanced. In general, an end-to-end neural text-to-speech (TTS) model consists of an acoustic model and a vocoder. The acoustic model converts the input text into a spectrogram in an autoregressive \cite{Tacotron, Tacotron2, DCTTS, TransformerTTS} or non-autoregressive \cite{FastSpeech, FastSpeech2, AlignTTS, ParaNet, VaraTTS} way. The vocoder converts the spectrogram into a waveform by traditional algorithms \cite{GriffnLim} or neural networks \cite{WaveNet, ParallelWaveNet, WaveGlow, ParallelWaveGAN, VocGAN, HiFiGAN}.

The modeling of non-linguistic attributes is important for synthesizing natural and expressive speech. Many TTS models learn and control style attributes, such as speaker ID \cite{Deepvoice2, ProsodyTacotron}, prosody \cite{ProsodyTacotron, CHIVE}, emotion \cite{EndtoEndEmotionalGST, ControllingEmotionalExpressiveness}, and acoustic features \cite{FastSpeech2}. They synthesize speech signals conditioned on embedding vectors that represent one or more attributes. To learn correlated attributes together, recently developed models represent the relationship between attributes by hierarchical methods \cite{ControllingEmotionalExpressiveness, FullyHierarchicalFineGrained, HierarchicalProsodyModeling, GMVAETacotron, HGST}. They learn the dependency between the prosodies at the levels of phonemes, syllables, words, and sentences \cite{ControllingEmotionalExpressiveness, FullyHierarchicalFineGrained, HierarchicalProsodyModeling, GMVAETacotron}, or the dependency between different types of prosodies \cite{FullyHierarchicalFineGrained, GMVAETacotron, HGST} using RNNs \cite{ControllingEmotionalExpressiveness}, hierarchically extended variational auto-encoders (VAE) \cite{FullyHierarchicalFineGrained, GMVAETacotron}, or by stacking network modules \cite{HierarchicalProsodyModeling, HGST}.

While most existing models represent multiple non-linguistic attributes either independently \cite{Deepvoice2, ProsodyTacotron, CHIVE, EndtoEndEmotionalGST, GSTTacotron, RobustandFineGraned} or hierarchically \cite{ControllingEmotionalExpressiveness, FullyHierarchicalFineGrained, HierarchicalProsodyModeling, GMVAETacotron, HGST}, certain attributes are correlated in a non-hierarchical way. For example, speaker ID affects many low-level prosodic attributes, such as timbre, pitch, energy, and speaking rate. When we control speaker ID together with pitch, we should consider the effect of the speaker ID on the pitch. Moreover, the prosodic attributes are also affected by emotion. In this sense, the effect of speaker ID overlaps with that of emotion, while the relationship between the two attributes is not completely hierarchical. Simply adding their embeddings to the phoneme representations can redundantly affect the prosodic attributes, and as a result, degrade fidelity or controllability. Despite the efforts of researchers \cite{DANN, DisentangledSpeakerandNuiance, MultispeakerEmotionalSpeechSynthesis, SpeechSPlit}, separating the effects of overlapping attributes in speech synthesis remains a challenging problem.

To avoid interference between overlapping attributes, we represent multiple style attributes in a unified embedding space. The proposed model, UniTTS, represents phonemes by absolute coordinates and style attributes by the residuals between the phoneme embeddings before and after applying the attributes. UniTTS predicts the residual embeddings of style attributes with a collection of residual encoders, each of which predict the embedding of a style attribute normalized by the means of the previously applied attributes. As a result, UniTTS minimizes redundancy between attribute embeddings. The residual encoders are trained by a novel knowledge distillation technique. In addition, we present a novel data augmentation technique inspired by the transforming autoencoder \cite{TransformingAutoEncoders} that improves fidelity and controllability over style attributes leveraging the unified embedding space.

UniTTS is based on FastSpeech2 \cite{FastSpeech2} and includes several improvements: First, UniTTS synthesizes high-fidelity speech while controlling both speaker ID and emotion without interference. Second, UniTTS can mix the styles of multiple speakers. Third, UniTTS has a fine-grained prosody model that improves output quality. Fourth, UniTTS is effective in synthesizing expressive speech, as it predicts prosodic attributes conditioned on the previously applied attributes including the speaker ID and emotion. Fifth, UniTTS provides a simple and effective way to control pitch and energy. Additionally, UniTTS provides a convenient way to separate the attribute embeddings from the phoneme representation simply by element-wise subtraction. 

In experiments, the visualization results exhibited that the proposed methods successfully learned the embeddings of phonemes, speaker ID, emotions, and other prosodic attributes. UniTTS produced high-fidelity speech signals controlling speaker ID, emotion, pitch, and energy. The audio samples synthesized by UniTTS are presented at \url{https://anonymous-authors2022.github.io/paper_works/UniTTS/demos/}. The main contribution of our work includes the followings:

\begin{itemize}
    \item We present a novel method to represent multiple style attributes in a unified embedding space together with phonemes using a collection of residual encoders.
    \item We present a novel method to learn the residual encoders that combines the knowledge distillation and normalization techniques.
    \item We present a novel TTS model, UniTTS, that synthesizes high-fidelity speech controlling speaker ID, emotion, and other prosodic attributes without interference.
    \item We present a novel data augmentation technique inspired by the transforming autoencoder that improves fidelity and controllability over prosodic attributes.
    \item We visualize the unified embedding space demonstrating that the proposed method effectively learns the representation of multiple attributes together with phoneme embeddings.
\end{itemize}

The remaining parts of this paper are organized as follows: Section 2 presents the background of our research. Section 3 introduces the unified embedding space and Section 4 explains the structure and learning algorithm of UniTTS. The experimental results and conclusions are presented in Sections 5 and 6, respectively.

\section{Background}
The acoustic model of a speech synthesizer converts the input text $y = (y_1, y_2, ... , y_L)$ into the corresponding spectrogram $x = (x_1, x_2, ... , x_T)$, where $y_i$ denotes a phoneme while $x_j$ denotes a frame of spectrogram. The model learns the conditional probability $P(x | y)$ from the training samples. The autoregressive TTS models produce one or more frames at each time-step from the input text and the previously synthesized frames according to the recursive formula $P(x | y) = \prod_{t=1}^T P(x_t | x_{<t}, y)$. On the other hand, the non-autoregressive TTS models predict the phoneme duration and align the text to the spectrogram by duplicating phoneme embeddings for the predicted durations\cite{FastSpeech, FastSpeech2, AlignTTS}. Then, they synthesize $x$ from $P(x | \tilde{y})$ parallelly using a feed-forward network, where $\tilde{y}$ is the expanded phoneme sequence whose length is the same as that of the output spectrogram. On the other hand, \cite{ParaNet} and \cite{VaraTTS} align the text and spectrogram by iteratively refining the initial alignment in a layer-by-layer way.

To control the style of the output speech, the TTS model produces speech signal conditioned on the embedding vectors of the style attributes. Such a model learns $P(x | y, z_1, z_2, …, z_N)$, where $z_k$ denotes a global attribute or a sequence of local attributes. When the attributes are assumed independent, $z_k$s can be represented in separate embedding spaces. For example, \cite{ProsodyTacotron} and \cite{GSTTacotron} represent speaker ID and unlabeled prosody as separate embedding vectors. Such models represent the attributes in separate embedding spaces as Fig. 1(a). However, although prior work has proposed disentangling methods, such as gradient reversal \cite{DANN} and information bottleneck \cite{SpeechSPlit}, it is still challenging to separate correlated attributes cleanly.

    \begin{figure}
         \centering
         \begin{subfigure}[b]{0.32\textwidth}
             \centering
             \includegraphics[width=3.5cm, height=3.5cm]{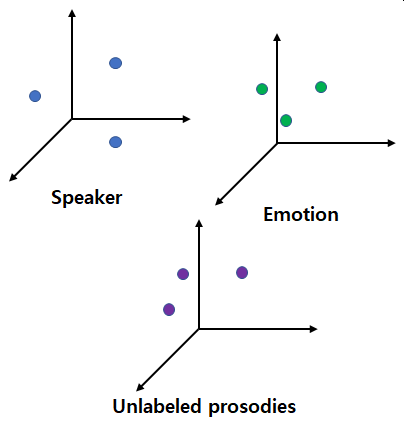}
             \caption{Separate embedding spaces}
         \end{subfigure}
         \hfill
         \begin{subfigure}[b]{0.32\textwidth}
             \centering
             \includegraphics[width=3.5cm, height=3.5cm]{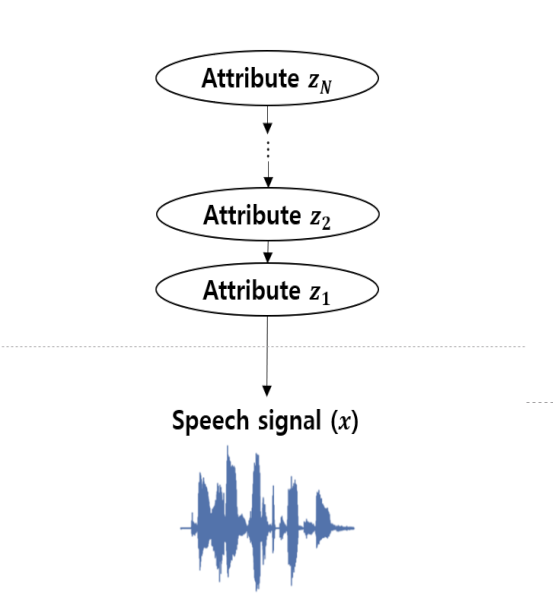}
             \caption{Hierarchical embedding spaces}
             \label{fig:three sin x}
         \end{subfigure}
         \hfill
         \begin{subfigure}[b]{0.32\textwidth}
             \centering
             \includegraphics[width=3.5cm, height=3.5cm]{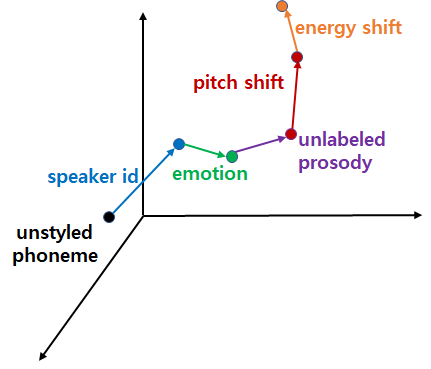}
             \caption{Unified embedding space}
             \label{fig:five over x}
         \end{subfigure}
         \caption{Embedding spaces to represent multiple style attributes.
         }
         \label{figure:embedding_spaces}
         
    \end{figure}

To learn correlated attributes, prior work has developed hierarchical representation, as shown in Fig. 1(b). \cite{VaraTTS, FullyHierarchicalFineGrained, HierarchicalProsodyModeling} have proposed hierarchical models based on hierarchically extended VAEs. For a training sample $(x, z_1, z_2, …, z_N)$, where $z_k \sim p(z_k | z_{k+1})$ and $x \sim p(x | z_1)$, the hierarchical VAE learns the conditional distribution $p(x | y, z_1, …, z_N)$ by maximizing the evidence lower bound using a series of approximate posteriors $q(z_k | x, z_{<k})$ as equation (\ref{eq:ELBO}), where $\ q(z_{<k} | x) = \prod_{i=1}^{k-1} q(z_i | x, z_{<i})$.
    \begin{equation}
        \begin{split}
                \log p(x) \ge \mathop{\mathbb{E}}{}_{z \sim  q(z|x)} \log p(x|z) & - KL[q(z_1|x) || p(z_1)] \\
                     & - \sum_{k=1}^N \mathop{\mathbb{E}}{}_{q(z_{<k} | x)} [KL[q(z_k  | z_{<k}, x] || p(z_k | z_{<k}))]
        \end{split}
        \label{eq:ELBO}
    \end{equation}
Certain prior works learn hierarchical representation by combining VAEs and GMMs \cite{HierarchicalProsodyModeling} or by stacking multiple network modules \cite{GMVAETacotron, HGST}. In \cite{GriffnLim,FullyHierarchicalFineGrained, HierarchicalProsodyModeling}, each of $z_k$s corresponds to phoneme-, word-, or utterance-level prosody, respectively. On the other hand, in \cite{FullyHierarchicalFineGrained, HierarchicalProsodyModeling, GMVAETacotron, HGST}, $z_1$ learns the low-level prosodic attributes, such as pitch, energy, and speaking speed, while $z_2$ and $z_3$ learn high-level attributes, such as speaker ID.

\section{Unified Embedding Space for Learning Multiple Style Attributes}

The motivation of the unified embedding space is to learn and control overlapping attributes avoiding interference. For example, if a bright-tone speaker and a calm-tone speaker speak in their normal tone, respectively, their utterances are different in speaker ID, while the same difference can be also interpreted as the difference in emotion. If speaker ID and emotion are represented in separate embedding spaces, it is not easy to deal with such overlap. One possible way to represent such overlapping attributes is to learn multiple attributes in a single embedding space. 
We represent attributes by the residuals between the phoneme embeddings before and after applying the attributes as Fig. 1(c).

In UniTTS, the phoneme encoder takes a sequence of phonemes as input and produces a sequence of high-level phoneme representations. We call each of them unstyled phoneme embedding and denote it as $E(y_i)$, as the style attributes have not been added, yet. Applying an attribute $z$ to $y_i$ moves the phoneme embedding to another coordinate, $E(y_i, z)$. As the two phoneme embeddings are in the same vector space, we can represent the effect of $z$ on $y_i$ by the residual between the phoneme embeddings before and after applying $z$ computed as $R(z | y_i) = E(y_i, z) - E(y_i)$.

When multiple attributes $z_1, …, z_N$ are applied to $y_i$ sequentially, the phoneme embedding moves following the path $E(y_i)$, $E(y_i, z_1)$, $E(y_i, z_1, z_2)$, ..., $E(y_i, z_1, ..., z_N)$. The embedding after applying attributes $z_1, ..., z_k$ is computed recursively by the sum of the previous embedding and the residual vector for $z_k$ as $E(y_i, z_1,..., z_k)=E(y_i, z_1, ..., z_{k-1})+R(z_k | y_i, z_{<k})$. In this case, $R(z_k | y_i, z_{<k})$ represents the effect of $z_k$ on $y_i$ conditioned on the previously applied attributes $z_1, ..., z_{k-1}$.
The phoneme embedding after applying all attributes $z_1, …, z_N$ is computed as $E(y_i, z_1,..., z_N) = E(y_i) + \sum_{k=1}^N R(z_k | y_i, z_{<k})$.

    \begin{figure}[h!]
    
        \medskip
        \centerline{\includegraphics[width=130mm]{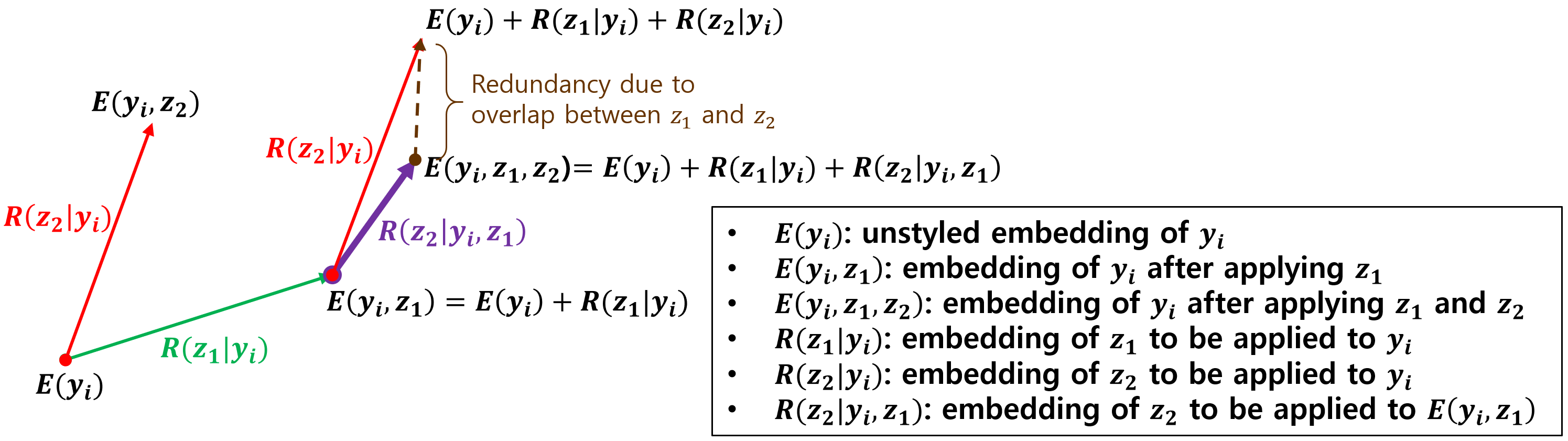}}
        \caption{Two overlapping attributes applied to a phoneme in the unified embedding space. UniTTS avoids redundancy caused by overlap between attributes by applying each attribute conditioned on the previously applied attributes. }
        \label{figure:residual_overlap}    
    \end{figure}

Fig. \ref{figure:residual_overlap} illustrates how the proposed method avoids redundancy when applying overlapping attributes. When $z_1$ and $z_2$ are applied to the unstyled embedding $E(y_i)$ independently, their residual vectors are $R(z_1|y_i)=E(y_i,z_1)-E(y_i)$ and $R(z_2|y_i)=E(y_i,z_2)-E(y_i)$, respectively. Adding the two attribute embeddings to $E(y_i)$ results in $E(y_i)+R(z_1|y_i)+R(z_2|y_i)$. Such a result can be different from $E(y_i, z_1, z_2)$, the actual embedding of $y_i$ after applying $z_1$ and $z_2$, because adding both residual vectors reflects the overlapping portion of their effects redundantly. In this case, the distance between $E(y_i)+R(z_1|y_i)+R(z_2|y_i)$ and $E(y_i, z_1, z_2)$ represents the amount of overlap between the effects of $z_1$ and $z_2$ on $y_i$. On the other hand, UniTTS applies $z_2$ to $E(y_i, z_1)$ by adding $R(z_2|y_i, z_1)$ instead of $R(z_2|y_i)$. Since $R(z_2|y_i, z_1)$ represents the effect of $z_2$ on $y_i$ conditioned on $z_1$ and $E(y_i, z_1)+R(z_k | y_i, z_1)=E(y_i, z_1,z_2)$ by definition, UniTTS does not reflect the overlapping attributes redundantly. We learn $R(z_k | y_i, z_{<k})$ with a residual encoder using a novel knowledge distillation technique. The following section explains the design and learning algorithm of UniTTS.

\section{High-Fidelity Speech Synthesis with Multiple Style Control}
\subsection{Model structure} 

\begin{figure}[h!]
        \centerline{\includegraphics[width=143mm]{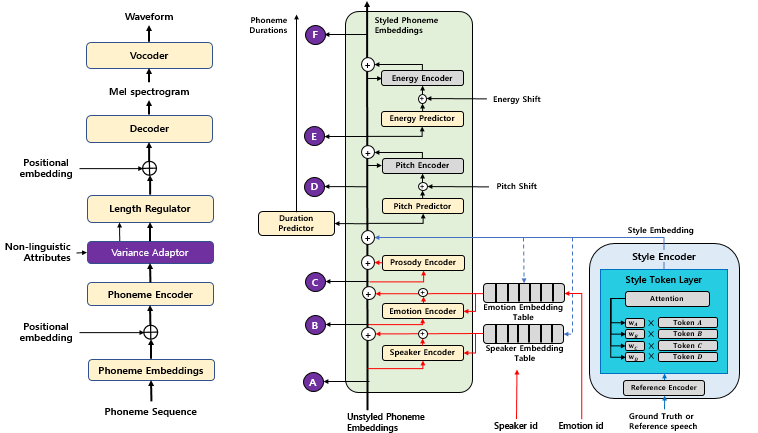}}
        \medskip
        \caption{The structure of UniTTS.}
        \label{figure:model_structure}    
    \end{figure}

The structure of UniTTS is based on FastSpeech2 \cite{FastSpeech2} and includes several improvements as illustrated in Fig. \ref{figure:model_structure}. The phoneme encoder extracts high-level representations from the input phonemes, and the variance adapter adds non-linguistic attributes. The length regulator expands the phoneme sequence by duplicating the phoneme representations for their durations. The decoder converts the expanded phoneme sequence into a Mel spectrogram, from which the vocoder synthesizes the waveform.

To synthesize high-fidelity expressive speech controlling multiple attributes, we extended the baseline variance adapter as follows: First, the length regulator was moved from inside the variance adapter to behind the variance adapter. This modification allows the variance adapter to process all information at the phoneme level. Additionally, prior work has shown that predicting variances at the phoneme level rather than at the frame level improves speech quality \cite{FastPitch}. Second, we introduced speaker and emotion encoders to add the variance in speaker ID and emotion. Based on the unified embedding space, UniTTS learns and controls the overlapping attributes without interference. Third, we extended the pitch and energy predictors to predict and encode pitch and energy conditioned on the previously applied attributes. Fourth, we added a style encoder to transfer style from a reference speech. More importantly, we distill the knowledge of the style encoder to learn other residual encoders. 

\paragraph{Style encoder}  When a reference speech is provided, the style encoder extracts a style embedding that carries non-linguistic attributes not included in the input text. The style encoder comprises a reference encoder and a style token layer, following \cite{GSTTacotron}. To produce a style embedding, the style token layer combines the token vectors through the multi-head attention \cite{Transformer}. The style encoder retrieves speech style not using speaker nor emotion label. Therefore, the style embedding is a mixture of various types of style attributes not separated by speaker or emotion. As the style of the reference speech and the variance in speaker ID and emotion overlap, we activate either the style encoder (blue lines in Fig. \ref{figure:model_structure}) or a collection of the speaker, emotion, and prosody encoders (red lines in Fig. \ref{figure:model_structure}), but not both. We train the style encoder using the ground truth speech samples as reference speech by the learning algorithm described in \cite{GSTTacotron}.

\paragraph{Speaker, emotion, and prosody encoders}  When the speaker and emotion IDs are specified, UniTTS adds the variance by the speaker and emotion IDs to the phoneme embeddings using the speaker and emotion encoders. As described in the previous section, simply adding the embeddings for the speaker and emotion IDs to the phoneme embeddings independently reflects the overlapping portion of the two attributes redundantly. One possible way to avoid such a problem is letting each embedding represent a single speaker-emotion pair. However, this requires training samples of all combinations of speaker IDs and emotion types. Combining their embeddings by a neural network as in \cite{MultiConditionEmotionalSpeechSynthesizer, MultispeakerEmotionalSpeechSynthesis} is more scalable. However, in such a network, the backpropagation algorithm cannot separate the gradients for the embedding of each attribute completely. Separating multiple attributes in a speech signal is still challenging, although some researchers are making progress \cite{MultispeakerEmotionalSpeechSynthesis, SpeechSPlit}.

As explained in the previous section, UniTTS avoids redundancy as it applies $z_k$ by adding $R(z_k | y_i, z_{<k})$ in which the overlap with the previously applied attributes is removed. In this research, we set speaker ID, emotion, and phoneme-level unlabeled prosody as $z_1$, $z_2$ and $z_3$, respectively. The residual vector for a speaker ID, $R(z_1 | y_i)$, is computed using an embedding table and an encoder that adapts the chosen embedding to phoneme $y_i$. The embedding table is learned by distilling the knowledge of the style encoder. As mentioned above, the style encoder produces the style embedding which is a mixture of various types of style attributes. We set the entry of the embedding table for a speaker $s_u$ to the mean of the style embeddings extracted from the samples spoken by $s_u$ as $\mu_{s_u} = \frac {1}{N_{s_u}} \sum_{u'=u}S(x_{u'v'})$,
where $\mu_{s_u}$ is the entry of the embedding table for $s_u$, and $S(\cdot)$ is the style encoder. $x_{uv}$ denotes a speech sample with a speaker label $s_u$ and an emotion label $e_v$. $N_{s_u}$ is the number of samples spoken by $s_u$. On the other hand, the speaker encoder takes as input $\mu_{s_u}$ and the phoneme embedding $E(y_i)$ and adapts $\mu_{s_u}$ to $E(y_i)$.

Similarly, the residual vector $R(z_2 | y_i, z_1)$ for an emotion type $e_v$ is computed using an embedding table and a residual encoder. However, when training the emotion embedding table, we normalize the style embedding $S(x_{uv})$ by subtracting $\mu_{s_u}$ to remove the overlap with the variance by the speaker ID as $\mu_{e_v} = \frac {1}{N_{e_v}} \sum_{v'=v} [S(x_{u'v'}) - \mu_{s_u}]$, 
where $\mu_{e_v}$ is the entry of the embedding table for an emotion type $e_v$ and $N_{e_v}$ is the number of samples with emotion label $e_v$. The visualization results in the next section demonstrate the proposed normalization can successfully learn the residual embeddings to represent emotion. As well, prior work has shown that such normalization improves robustness \cite{RobustandFineGraned}.

With a pretrained style encoder, we first learn the embedding tables by the knowledge distillation, and then, learn the speaker and emotion encoders freezing the embedding tables. The proposed distillation method allows UniTTS to minimize  redundancy, and additionally, helps the speaker and emotion encoders to learn quickly.

Additionally, UniTTS includes a fine-grained prosody model that learns unlabeled prosody at the phoneme level. Prior work has shown that the fine-grained prosody model improves speech quality \cite{HierarchicalProsodyModeling}. One way to learn the residual embedding of unlabeled prosody not reflected by the speaker and emotion IDs is the aforementioned distillation technique to learn the style embedding normalized by the speaker and emotion embeddings as $S(x_{uv}) - \mu_{s_u} - \mu_{e_v}$. However, we attempted a simpler trick and it worked fine. We first learn the speaker and emotion encoders, and then learn the phoneme level prosody model freezing the speaker and emotion encoders. The residual prosody encoder predicts the residual $R(z_3|y_i, z_1, z_2) = E(y_i, z_1, z_2, z_3) -  E(y_i, z_1, z_2)$, where $z_1, z_2, z_3$ are speaker ID, emotion, and prosody, respectively. Freezing the speaker and emotion encoders fixes $E(y_i, z_1, z_2)$. The output of the residual prosody encoder is added to $E(y_i, z_1, z_2)$ to compute $E(y_i, z_1, z_2, z_3)$ as $E(y_i, z_1, z_2, z_3) = E(y_i, z_1, z_2) + R(z_3|y_i, z_1, z_2)$. When the learning algorithm optimizes $E(y_i, z_1, z_2, z_3)$, fixing $E(y_i, z_1, z_2)$ forces the model to focus on $R(z_3 | y_i, z_1, z_2)$.


\paragraph{Duration, pitch, and energy predictors}  The duration predictor predicts phoneme durations to provide to the length regulator. The pitch and energy predictors add the variance of pitch and energy to the phoneme embeddings. The variance of duration is reflected by the length regulator, while pitch and energy are modeled as $z_4$ and $z_5$, respectively. We extended the predictors of FastSpeech2 to predict the variances based on the previously applied attributes, e.g., the speaker and emotion IDs. While the predictors of FastSpeech2 take an unstyled phoneme embedding $E(y_i)$ as input, those of UniTTS take as input $E(y_i, z_{<k})$. As a result, they predict duration, pitch, and energy conditioned on both the grapheme sequence and the previously applied attributes. For example, our duration predictor can predict the duration of the same phoneme differently according to the speaker and emotion IDs. 

While the pitch (energy) predictor of FastSpeech2 outputs the pitch (energy) embedding chosen from the embedding table, we predict the pitch (energy) using a residual encoder separated from the pitch (energy) predictor. While the embedding table consists of fixed vectors for each predicted value, our residual encoder can output embeddings adapted to the input phoneme embeddings, and therefore, more appropriate to implement the idea of the residual learning described in the previous section.

Since the encoder was separated from the predictor, we can manually adjust pitch and energy by adding the desired shift to the predicted pitch and energy values. This draws an additional advantage that we can apply a data augmentation technique to improve the learning of the predictors and encoders. Training the predictors and encoders to control the pitch and energy sufficiently requires speech samples with various pitch and energy values. To increase variety in pitch and energy, we applied a data augmentation technique inspired by the transforming autoencoder \cite{TransformingAutoEncoders}. We generated augmented samples by adjusting the pitch and energy of the training samples using a off-the-shelf speech processing toolkit, Sound of eXchange (SoX) \cite{SoX}. The amounts of pitch and energy shift for each sample were randomly selected from a pre-determined range (pitch: [-400,400] cents, energy: [0.3,1.7]). Then, we trained UniTTS with both the original and augmented training samples. When we trained with an augmented sample, we fed the amount of pitch/energy shift as the ‘Pitch Shift’ and ‘Energy Shift’ in Fig. 3. We added the pitch shift to the predicted pitch value (the output of the pitch predictor) and multiply the energy shift to the predicted energy value (the output of the energy predictor). This informs the model of the pitch and energy shift of the augmented training sample, so that the model can learn pronunciation and style without being confused by the change in pitch or energy.

\subsection{Separating and visualizing attribute embeddings}  

Composed of a fully residual structure, the variance adapter of UniTTS combines the embeddings of style attributes to the phoneme embeddings by element-wise addition. This makes it easy to separate style attributes. In Fig. \ref{figure:model_structure}, the embedding vectors at the locations marked as $A$, $B$, ..., $F$ represent $E(y_i)$, $E(y_i,z_1)$, ..., $E(y_i,z_1,...,z_N)$, respectively. We can restore the residual embeddings of the sub-sequence of the attributes, $R(z_k,...,z_l|y_i, z_{<k})$ for any $k$ and $l$, $(k \le l)$, e.g., $R(z_1|y_i)=B-A$ and $R(z_2|y_i,z_1)=C-B$. $F-A$ corresponds to the full-style embedding $R(z_1, ..., z_N | y_i)$ that accumulates the variance of all attributes. On the other hand, $F-B$ contains all attributes but the speaker ID. We present the visualization results of the residual embeddings in the next section.

\section{Experiments}
\subsection{Experimental settings} \label{experimental_settings}
We used three speech datasets in experiments: The Korean Single Speech (KSS) dataset \cite{KSS}, The Korean Emotional Speech (KES) dataset \cite{KoreanEmotionalSpeech}, and The EmotionTTS Open DB (ETOD) dataset \cite{EmotionTTS}. The KSS \cite{KSS} dataset contains 12,853 samples without emotion labels spoken by a single female speaker. The KES \cite{KoreanEmotionalSpeech} dataset contains 22,087 samples with 7 emotion types (neutral, happy, sad, angry, disgusting, fear, surprise) spoken by a single female speaker. The ETOD \cite{EmotionTTS} dataset contains speech samples with 4 emotion types (neutral, happy, sad, and angry) spoken by 15 speakers (8 males and 7 females). The number of samples per the combination of speaker and emotion is 100. The total number of samples in ETOD is 15 speakers * 4 emotion types * 100 samples = 6,000. Combining the three datasets, we used 41,706 samples with 7 emotion types spoken by 17 speakers. We used mel-spectrograms preprocessed using Han-window with filter length 1024, hop length 256, and window length 1024. We used speech samples with 22,050kHz sampling rate.

We built UniTTS based on the open source implementation of FastSpeech2 \cite{FastSpeech2_opensource}. The detail of the model structure, hyper-parameters, and training methods are presented in the appendix. We ran the experiments on a computer equipped with a Xeon E5-2630 v4 CPU and two NVIDIA GTX-1080Ti GPUs. The learning requires about one day when data augmentation was not applied, and about 4 days when applied. In MOS test, we asked 12 subjects to evaluate the fidelity and the similarities with the ground truth data in each style attributes. Although the number of subjects may seem small, many previous papers on speech synthesis present MOS results measured from 4 to 20 raters, presumably due to the high cost of the MOS test on speech samples.

\subsection{Experimental results}

\subsubsection{Visualization of the unified embedding space}

\begin{figure}
         \centering
         \begin{subfigure}[b]{0.32\textwidth}
             \centering
             \includegraphics[width=3.5cm, height=3.5cm]{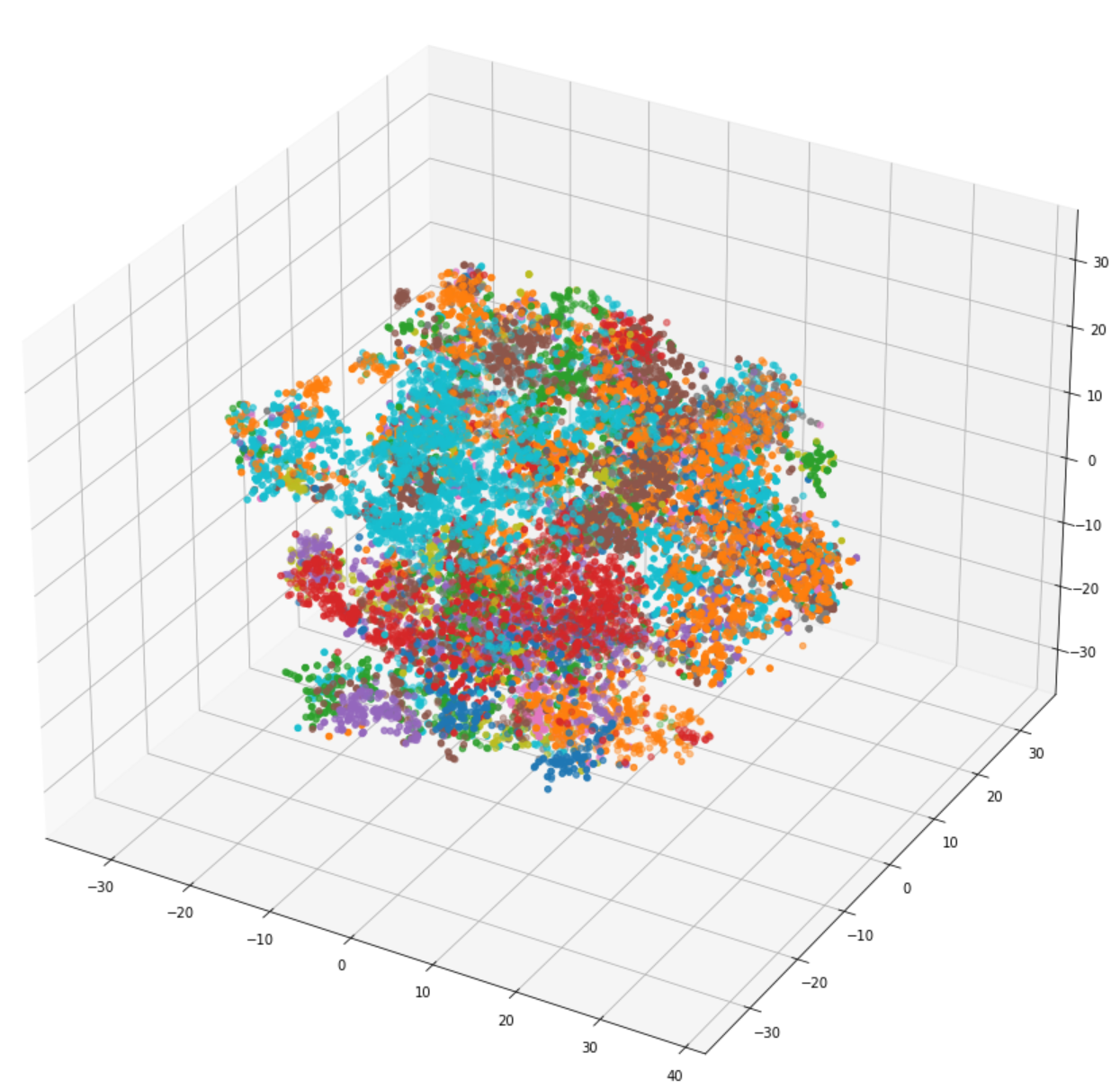}
             \caption{Unstyled phoneme embeddings colored by phoneme type}
         \end{subfigure}
         \begin{subfigure}[b]{0.32\textwidth}
             \centering
             \includegraphics[width=3.5cm, height=3.5cm]{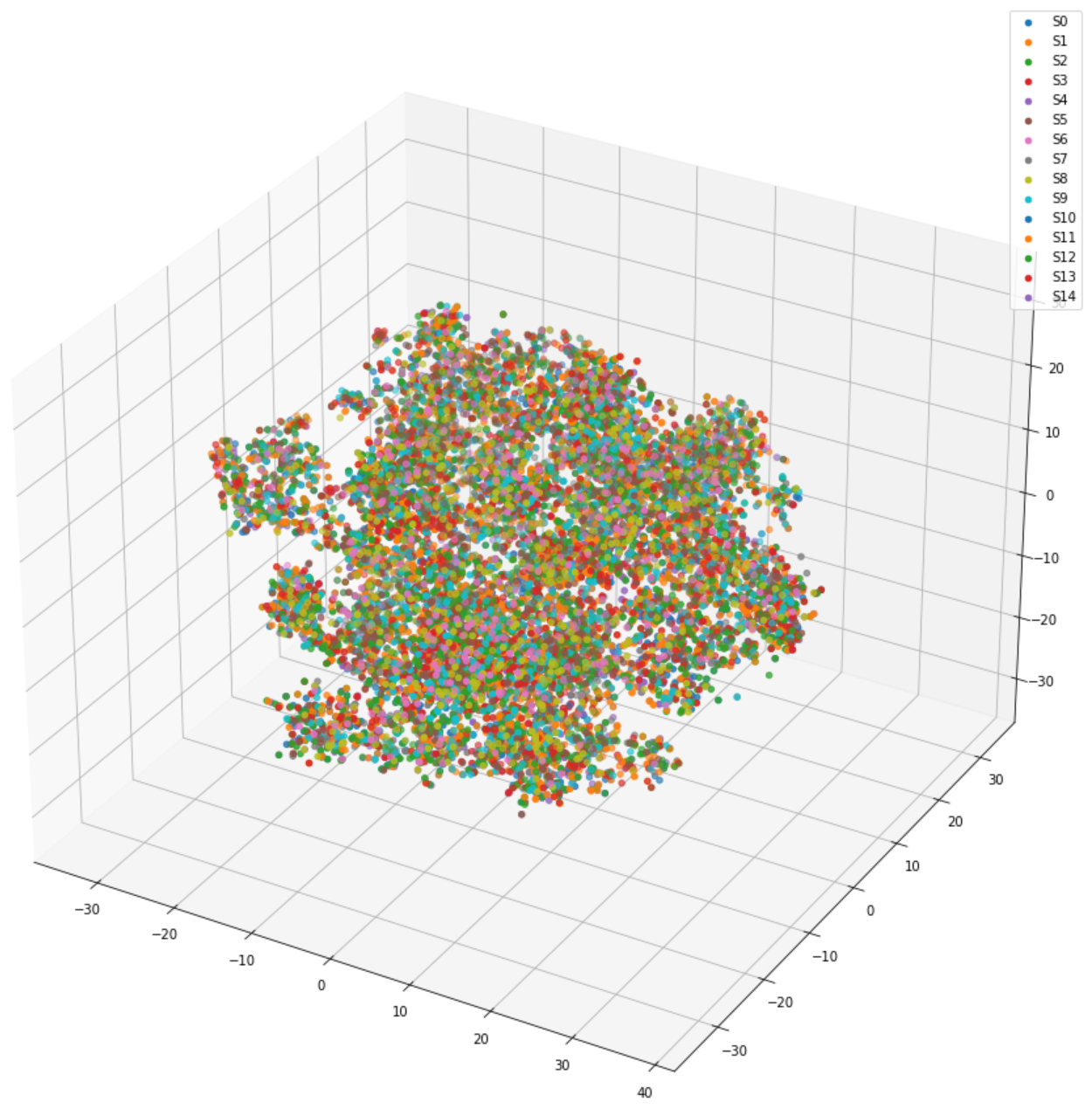}
             \caption{Unstyled phoneme embeddings colored by speaker label}
         \end{subfigure}
         \begin{subfigure}[b]{0.32\textwidth}
             \centering
             \includegraphics[width=3.5cm, height=3.5cm]{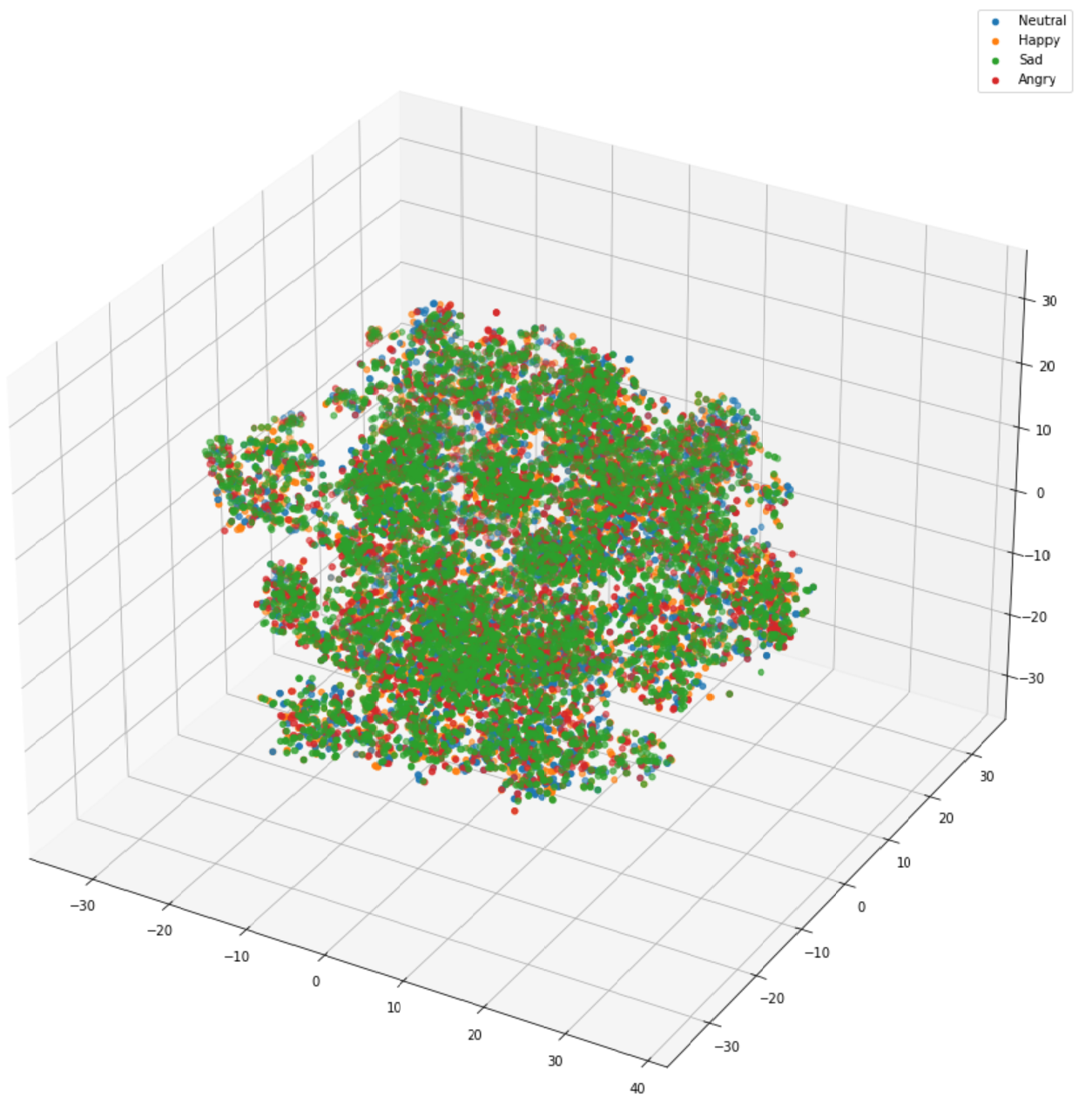}
             \caption{Unstyled phoneme embeddings colored by emotion label}
         \end{subfigure}
         \caption{The distribution of the unstyled phoneme embeddings extracted from the locations marked as A in Fig. \ref{figure:model_structure}. (a) shows that the unstyled phoneme embedding represents phoneme types, while (b) and (c) show that it does not contain speaker or emotion information.}
         \label{figure:unstyled_phoneme_embeddings}
         
    \end{figure}

We visualized the embeddings of the phonemes and style attributes learned by the proposed methods. We extracted the embeddings from the locations marked by the uppercase letters in Fig. \ref{figure:model_structure}. Fig. \ref{figure:unstyled_phoneme_embeddings}-F \ref{figure:full_style_embeddings} illustrate the distribution of embeddings by style attributes. The dots well-clustered according to the colors suggest that the embeddings are highly correlated with the attribute represented by the color (phoneme type, speaker ID, or emotion). Although UniSpeech learned 7 emotion types, we used the ETOD dataset that contains only 4 emotion types for visualization, because the KES dataset, that contains 7 emotion types, is a single-speaker dataset inappropriate for visualizing the distribution of embeddings by both speaker ID and emotion type.

\begin{figure}
         \centering
         \begin{subfigure}[b]{0.24\textwidth}
             \centering
             \includegraphics[width=\textwidth]{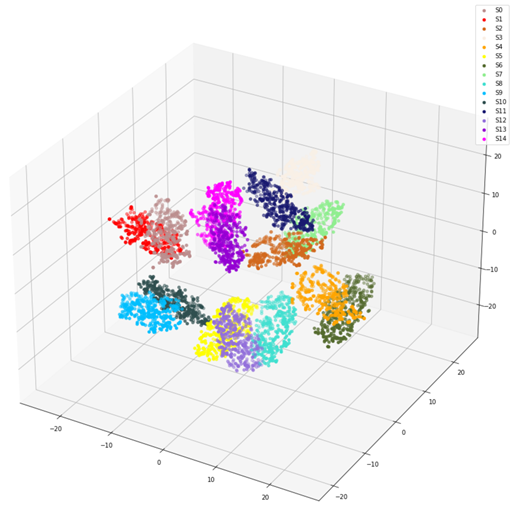}
             \caption{Speaker embeddings (B-A) colored by speaker label }
         \end{subfigure}
         \hfill
         \begin{subfigure}[b]{0.24\textwidth}
             \centering
             \includegraphics[width=\textwidth]{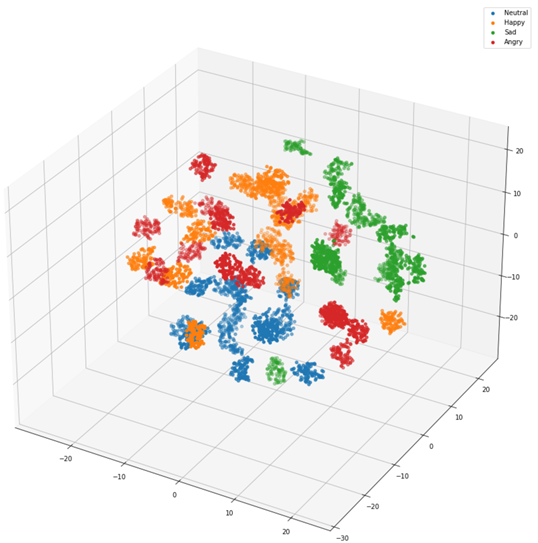}
             \caption{Emotion embeddings (C-B) colored by emotion label }
         \end{subfigure}
         \hfill
         \begin{subfigure}[b]{0.24\textwidth}
             \centering
             \includegraphics[width=\textwidth]{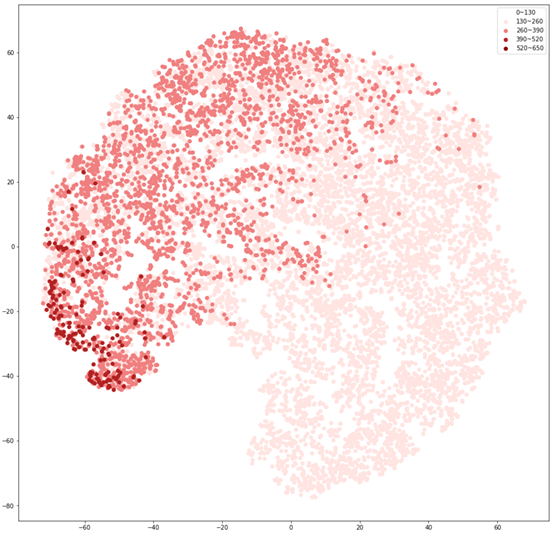}
             \caption{Pitch embeddings (E-D) colored by predicted pitch value}
         \end{subfigure}
         \hfill
         \begin{subfigure}[b]{0.24\textwidth}
             \centering
             \includegraphics[width=\textwidth]{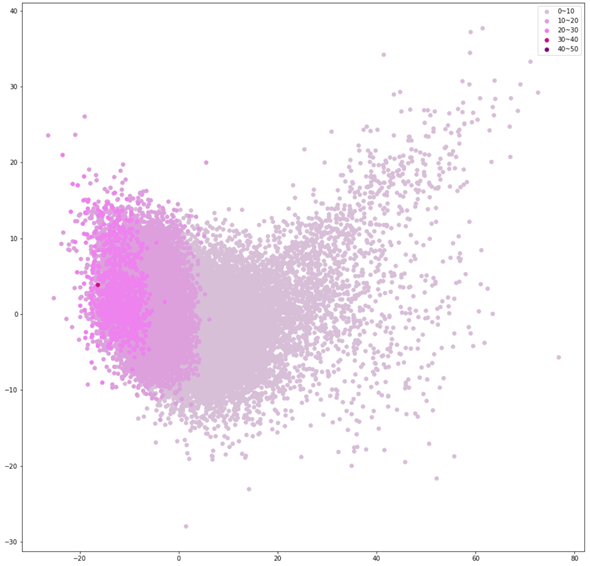}
             \caption{Energy embeddings (F-E) colored by predicted energy value}
         \end{subfigure}
         
         \caption{The distribution of the residual embeddings of speaker, emotion, pitch, and energy. The uppercase letters indicate the locations in Fig. \ref{figure:model_structure} where the embeddings were extracted. These figures show that the residual embeddings are effective in representing the style attributes.}
         \label{figure:individual_style_embeddings}
    \end{figure}

\begin{figure}
         \centering
         \begin{subfigure}[b]{0.24\textwidth}
             \centering
             \includegraphics[width=\textwidth]{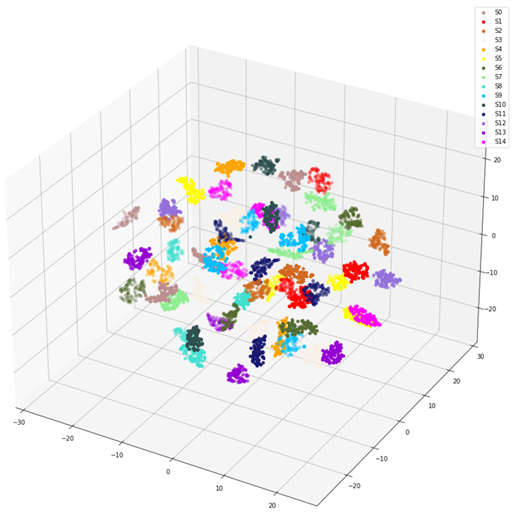}
             \caption{Full-style embeddings (F-A) colored by speaker label \\}
         \end{subfigure}
         \hfill
         \begin{subfigure}[b]{0.24\textwidth}
             \centering
             \includegraphics[width=\textwidth]{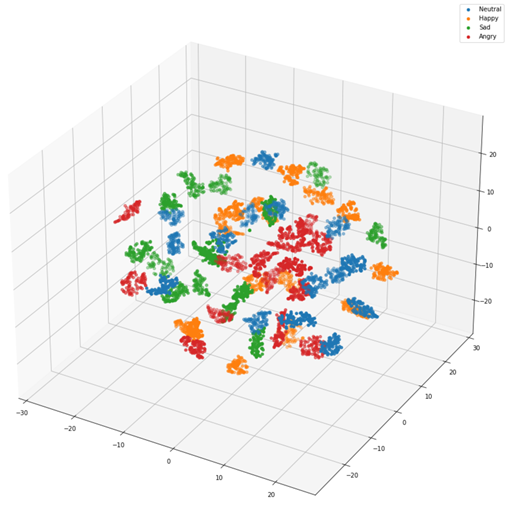}
             \caption{Full-style embeddings (F-A) colored by emotion label \\}
         \end{subfigure}
         \hfill
         \begin{subfigure}[b]{0.24\textwidth}
             \centering
             \includegraphics[width=\textwidth]{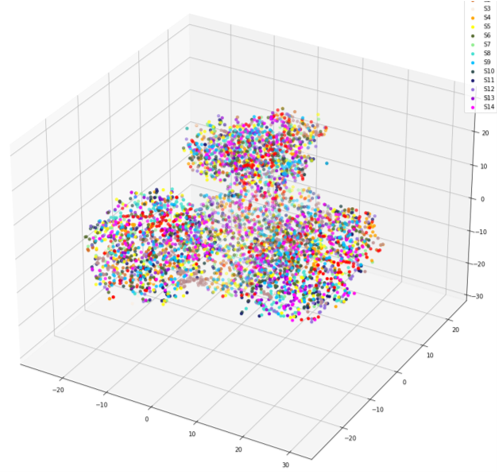}
             \caption{Full-style embeddings normalized by speaker embedding (F-B) colored by speaker label}
         \end{subfigure}
         \hfill
         \begin{subfigure}[b]{0.24\textwidth}
             \centering
             \includegraphics[width=\textwidth]{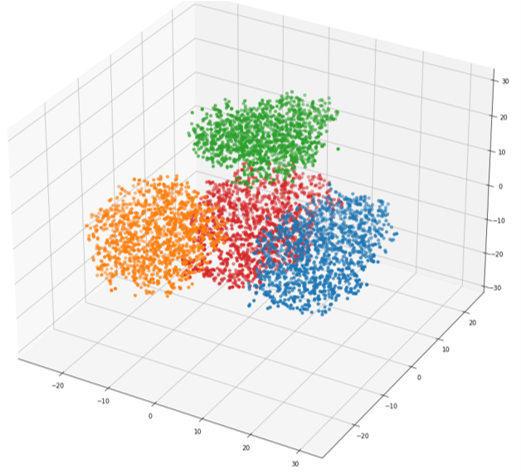}
             \caption{Full-style embeddings normalized by speaker embedding (F-B) colored by emotion label}
         \end{subfigure}
         \caption{The distribution of the full-style embeddings that incorporate all style attributes. The uppercase letters indicate the locations in Fig. \ref{figure:model_structure} where the embeddings were extracted. (a) and (b) show that the full-style embedding contains both speaker and emotion information. (c) shows that the full-style embedding normalized by the means of the speaker embeddings does not contain speaker information. (d) shows that the variance in emotion is dominant after normalizing the full-style embedding by the means of speaker embeddings.}
         \label{figure:full_style_embeddings}

    \end{figure}

Fig. \ref{figure:unstyled_phoneme_embeddings} displays the distribution of the unstyled phoneme embeddings. It suggests that the unstyled embedding carries the information about phoneme type, but not about speaker ID or emotion type. Fig. \ref{figure:individual_style_embeddings} exhibits the residual embeddings of speaker ID, emotion, pitch, and energy computed by the difference between the phoneme embeddings before and after the residual encoders as (B-A), (C-B), (E-D), and (F-E). The distributions well-clustered by colors clearly show that the residual embeddings of speaker ID, emotion type, pitch, and energy are closely correlated to the corresponding style attribute.
Fig. \ref{figure:full_style_embeddings} displays the distribution of the full-style embedding that incorporates all style attributes. The full-style embeddings were computed by the difference between the embeddings extracted from F and A. Fig. \ref{figure:full_style_embeddings}(a) and (b) show that the full-style embedding contains the variance in both speaker ID and emotion. However, after normalizing the full-style embedding by the mean of the speaker embeddings, the normalized full-style embedding does not contain the speaker information any more as (c), and as a result, the variance is mainly from emotion as shown in (d). Those visualization results strongly suggest that the proposed residual embeddings effectively represent multiple types of speech style attributes. 

\subsubsection{Fidelity and style control} \label{fidelity_and_style_control}

    \begin{table}[]
        \centering
        \caption{The results of the MOS test. UniTTS exhibited improved fidelity, speaker similarity, and emotion similarity compared with the other two models. In ablation study, the fidelity and emotion similarities were decreased when the data augmentation and the local prosody modeling were not applied. However, the speaker similarity was improved when the data augmentation was not applied. The reason is explained in Subsection \ref{fidelity_and_style_control}}
        
        \medskip
        \begin{tabular}{@{}ccccccc@{}}
        \toprule
        \textbf{Method}                                                       & \textbf{GT} & \textbf{\begin{tabular}[c]{@{}c@{}}GST\\ FS2\end{tabular}} & \textbf{\begin{tabular}[c]{@{}c@{}}Separate\\ Embeddings\end{tabular}} & \textbf{UniTTS} & \textbf{\begin{tabular}[c]{@{}c@{}}UniTTS\\ w/o data
        aug.\end{tabular}} & \textbf{\begin{tabular}[c]{@{}c@{}} UniTTS\\w/o local pros.\end{tabular}} \\ \midrule
        \textbf{Fidelity}                                                     & 4.88        & 3.30                                                       & 3.15                                                                   & \textbf{3.77}               & 3.61                                                             &    3.31                                                            \\ \midrule
        \textbf{\begin{tabular}[c]{@{}c@{}}Speaker\\ similarity\end{tabular}} & -           & 3.69                                                       & 3.73                                                                   & 3.88               & \textbf{3.96}                                                             & 3.73                                                               \\ \midrule
        \textbf{\begin{tabular}[c]{@{}c@{}}Emotion\\ similarity\end{tabular}} & -           & 3.31                                                       & 3.90                                                                   & \textbf{4.15}               & 3.98                                                             & 3.90                                                               \\ \bottomrule
        \end{tabular}
        \label{table:MOS_result}
    \end{table}

We synthesized speech by varying the speaker ID and emotion type. A few examples of the synthesized spectrograms are presented in the demo page. UniTTS produced speech signals with different styles according to the speaker ID and emotion type. We also ran an MOS test to evaluate fidelity and the ability to express the speaker characteristics and emotion. We compared UniTTS with the ground truth, denoted by 'GT' in Table \ref{table:MOS_result}, and two baseline models that can control both speaker ID and emotion type: The first model combines FastSpeech2 \cite{FastSpeech2} and a style encoder composed of a style token layer \cite{GSTTacotron}. With this model, we specify the desired speech style by providing a reference speech. The style encoder extracts a style embedding and adds it to the phoneme representations before the duration predictor. (Please refer to Fig. 1 of \cite{FastSpeech2}.) The style embedding guides the model to produce a speech signal of a style similar to that of the reference speech. This model exhibits high-fidelity but does not allow to control individual style attributes by directly feeding a speaker or emotion ID. To produce an speaker-specific output, we input a reference speech spoken by the target speaker. Similarly, we input a reference speech with the target emotion label to produce a emotion-specific output.  
We used this model as a teacher for learning UniTTS by distillation.
The other baseline model is composed of the same structure with UniTTS except the way to represent speaker ID and emotion: it learns the attributes using separate embedding spaces as Fig. \ref{figure:embedding_spaces}(a). Instead of the speaker and emotion encoders described above, this model has two embedding tables, one for speaker ID and the other for emotion, similar to \cite{MultispeakerEmotionalSpeechSynthesis} and \cite{MultiConditionEmotionalSpeechSynthesizer}. The two embedding tables are learned together with the other parts of the model. This model was designed to directly compare the proposed unified embedding space (Fig. \ref{figure:embedding_spaces}-(c)) with the conventional style representation method that uses separate embedding spaces (Fig. \ref{figure:embedding_spaces}-(a)). 

The results of the MOS test are presented in Table \ref{table:MOS_result}, where speaker similarity and emotional similarity are metrics to evaluate speaker characteristics and emotional expression performance of TTS models widely used in previous work on speech style modeling\cite{MultispeakerEmotionalSpeechSynthesis, MultiConditionEmotionalSpeechSynthesizer, Seq2seqEmotionalVoiceConversion}.
In Table \ref{table:MOS_result}, UniTTS exhibited higher MOS score than the other models in fidelity, speaker similarity, and emotion similarity.
Table \ref{table:MOS_result} also shows the results of the ablation study. When we did not apply the data augmentation and the phoneme-level local prosody modeling, the fidelity and emotion similarity were decreased. However, speaker similarity was slightly increased when we did not apply the data augmentation. Adjusting pitch by a software toolkit \cite{SoX} causes a side effect that changes the timbre of the voice. We believe such samples affected the learning of the TTS model negatively. We ran another MOS test to evaluate the effect of the data augmentation. This time, we asked the subjects to evaluate how much the pitch and volume of the synthesized samples are similar to those of the augmented training samples. The results are presented in Table \ref{Table:MOS_augmentation}. The proposed data augmentation improved the control over the pitch and energy significantly.



\begin{table}[]
    \centering
    \caption{The results of MOS test to evaluate the effect of data augmentation. The proposed data augmentation technique significantly improved the control over the pitch and energy of the synthesized speech.}
    \medskip

    \begin{tabular}{@{}ccc@{}}
    \toprule
    \multirow{2}{*}{\textbf{Method}} & \multicolumn{2}{c}{\textbf{Similarity to the augmented ground-truth}} \\ \cmidrule(l){2-3} 
                                     & \textbf{Pitch adj.}               & \textbf{Energy adj.}              \\ \midrule
    \textbf{UniTTS}               & \textbf{4.30}                     & \textbf{4.38}                     \\ \midrule
    \textbf{UniTTS w/o data aug.}           & 3.46                              & 3.33                              \\ \bottomrule
    \end{tabular}
    
    \label{Table:MOS_augmentation}
\end{table}

\subsubsection{Style mixing}

\begin{figure}[!htb]
        \centerline{\includegraphics[width=143mm]{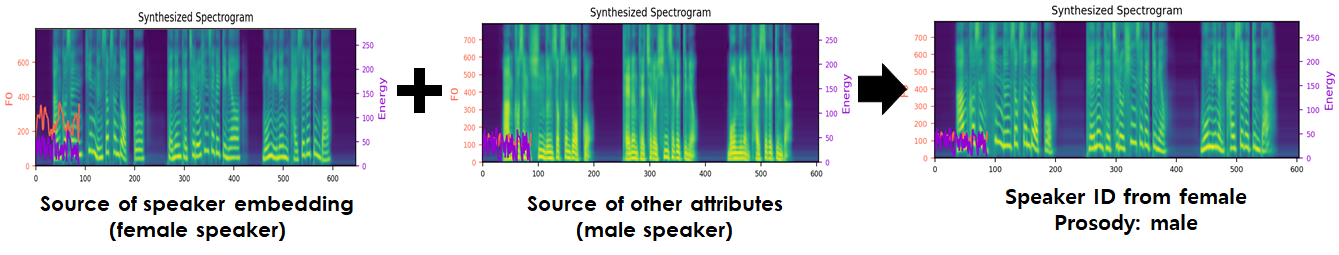}}
        \caption{Synthesizing speech by mixing the styles of different speakers. The third spectrogram was synthesized from the speaker embedding for the first spectrogram and the embeddings of the other style attributes for the second spectrogram. The synthesized speech combines the timbre of the female speaker and the style of the male speaker.}
        \label{figure:style_mixing}
    \end{figure}

Based on the unified embedding space, we synthesized speech by mixing the styles of different speakers. First, we synthesized speech twice using the IDs of two speakers saving the residual embeddings of all style attributes. Then, we synthesized a new speech sample mixing the saved embeddings. UniTTS successfully synthesized speech with the mixed style embeddings. The spectrogram of an example is presented in Fig. \ref{figure:style_mixing}. The audio samples are presented at the demo URL.

\section{Conclusion}

We proposed a novel expressive speech synthesizer, UniTTS, that synthesizes high-fidelity speech signals while controlling multiple non-linguistic attributes, such as speaker ID, emotion, duration, pitch, and energy. UniTTS represents non-linguistic attributes by residual vectors in a single unified embedding space. UniTTS can synthesize speech signals based on the specified speaker and emotion IDs or the style embedding extracted from a reference speech. UniTTS predicts prosodic attributes, such as phoneme duration, pitch, and energy, based on the speaker and emotion IDs since it predicts and encodes the embeddings of the prosodic attributes conditioned on the previously applied attributes. Additionally, we proposed a data augmentation technique to improve fidelity and controllability over attributes. The proposed method effectively learns and controls multiple overlapping attributes without interference.

\begin{ack}
    \begin{itemize}
    \item This work was supported by SkelterLabs, co., ltd.
    \item This work was supported by the National Program for Excellence in Software at Handong Global University (2017-0-00130) funded by the Ministry of Science and ICT.
    \item This study (or Project) used on open Speech database as the result of research supported by the Ministry of Trade, Industry \& Energy (MOTIE, Korea) under Industrial Technology Innovation Program (No. 10080667, Development of conversational speech synthesis technology to express emotion and personality of robots through sound source diversification).
    \end{itemize}
\end{ack}

\newpage
\bibliographystyle{achemso}
\bibliography{reference}

\newpage

\newpage
\appendix
\section*{Appendix}

\section{The detailed structures of the encoders and predictors}
        \begin{figure}[h]
            \begin{subfigure}[b]{0.27\textwidth}
                 \centering
                 \includegraphics[width=\textwidth]{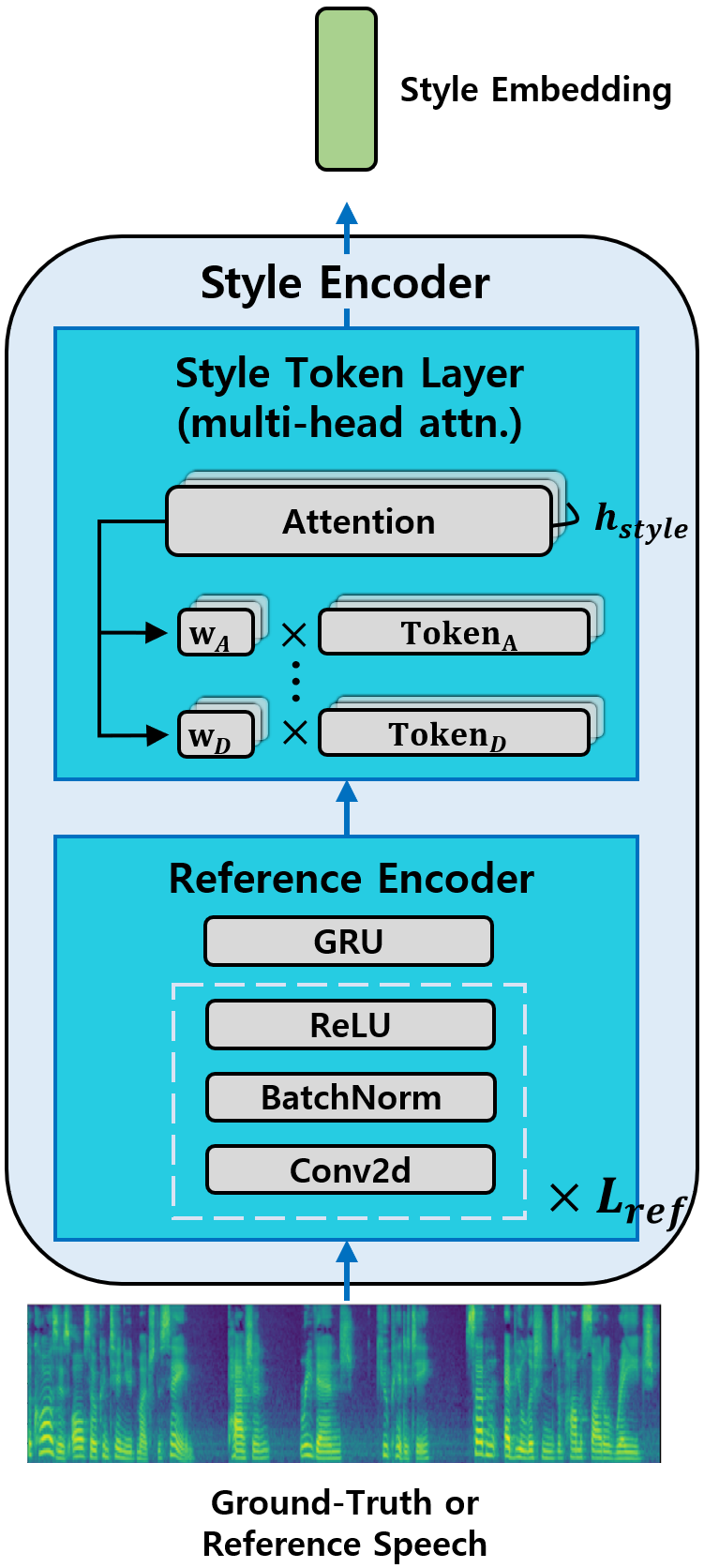}
                 \caption{Style encoder \newline }
            \end{subfigure}
            \hfill
            \begin{subfigure}[b]{0.2\textwidth}
                 \centering
                 \includegraphics[width=\textwidth]{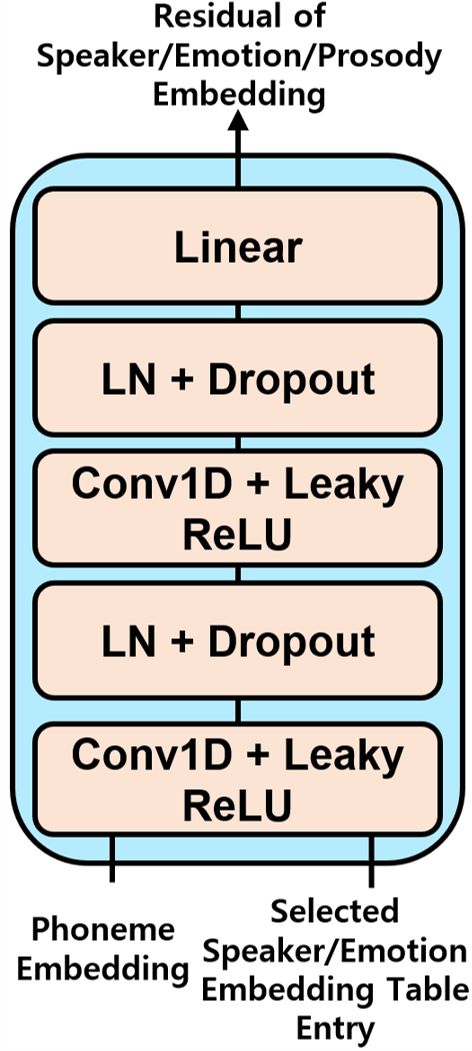}
                 \caption{Speaker/emotion/ prosody encoder }
             \end{subfigure}
             \hfill
            \begin{subfigure}[b]{0.195\textwidth}    
                 \centering
                 \includegraphics[width=\textwidth]{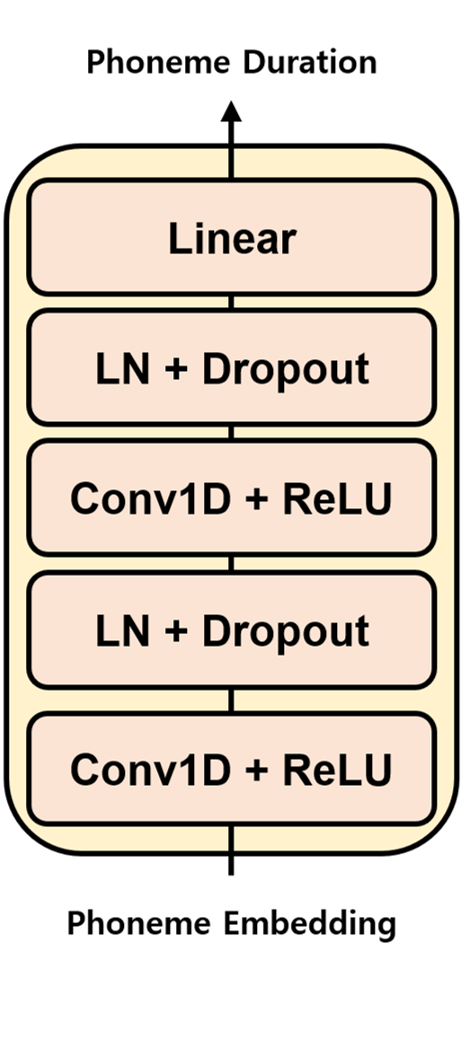}
                 \caption{Duration predictor \newline  }
             \end{subfigure}
             \hfill
             \begin{subfigure}[b]{0.31\textwidth}
                 \centering
                 \includegraphics[width=\textwidth]{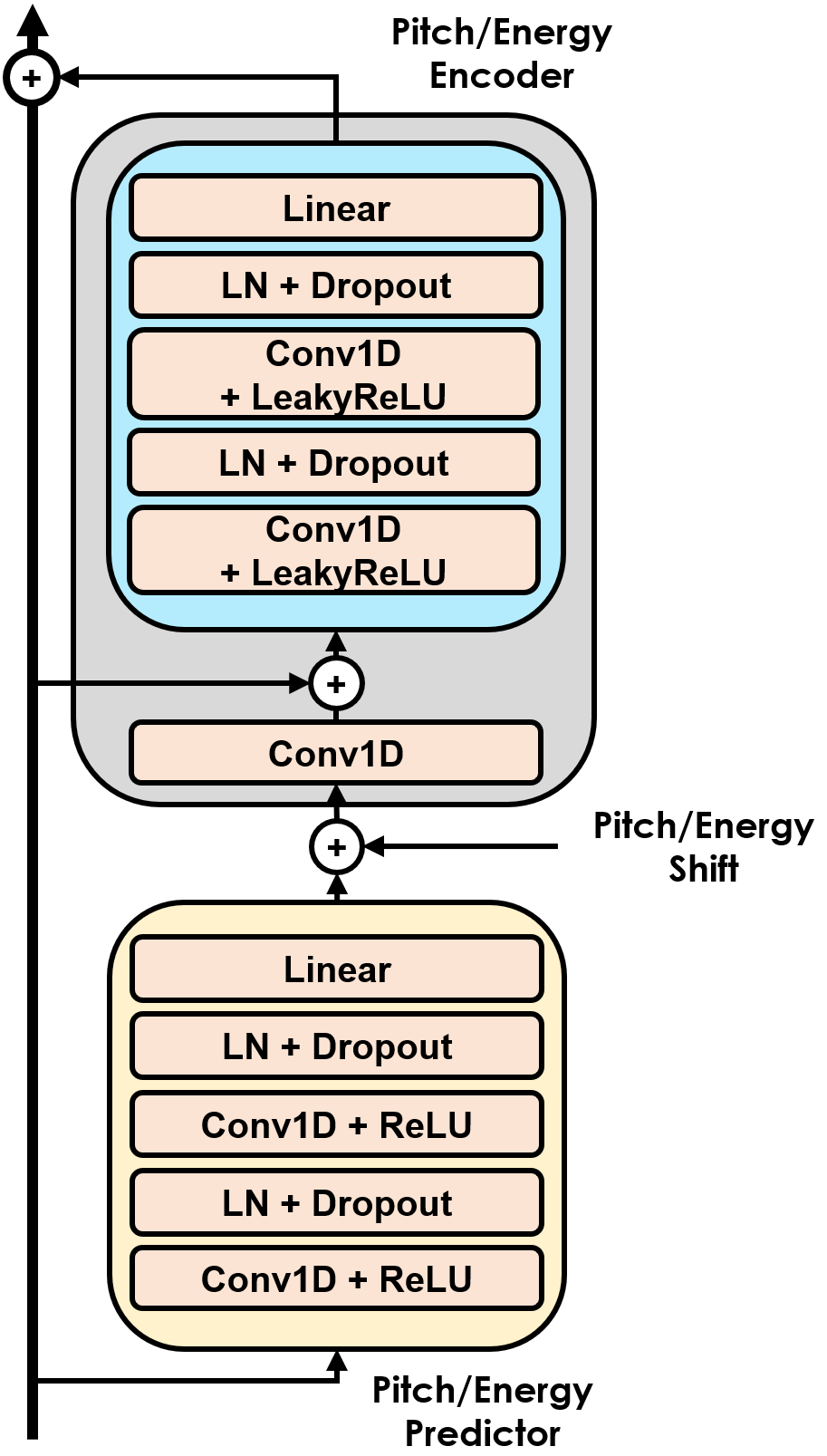}
                 \caption{Pitch/energy predictor \newline and encoder }
             \end{subfigure}             
     
            \caption{ The detailed structures of the predictors and encoders in Fig. \ref{figure:model_structure}. (a) The style encoder extracts a style embedding from a reference speech or ground-truth sample. It's structure is the same as that of \cite{GSTTacotron}, where $L_{ref}$ and $h_{style}$ are the number of Conv2d-BatchNorm-ReLU blocks and the number of attention heads, respectively. (b) The speaker/emotion encoder adapts the selected entry of the speaker/emotion embedding table by adding a residual vector. The prosody encoder outputs phone-level prosody embeddings. (c) The duration predictor predicts phoneme durations and pass them to the length regulator. (d) The pitch/energy predictor predicts the pitch/energy values of the phoneme embeddings. The pitch/energy encoder adds the pitch/energy embeddings to the phoneme embeddings. The proposed architecture allows adjusting pitch or energy by adding a pitch/energy shift to the predicted value. When we train the model with an augmented sample whose pitch or energy value was modified, we set the pitch or energy shift to the value used to augment the sample. This trick leads the model to learn to synthesize speech with a shifted pitch or energy value \cite{TransformingAutoEncoders}. }
            \label{fig:predictors_and_encoders}
        \end{figure}

    


\newpage

\section{Hyperparameters}
    
    \begin{table}[h]
        \centering
        \caption{The hyperparemeters of FastSpeech2 and UniTTS}
        \medskip
        
        \begin{tabular}{@{}cccc@{}}
        \toprule
        \multicolumn{2}{c}{\textbf{Hyperparameters}}                                                                                                                                    & \textbf{FastSpeech2} & \textbf{UniTTS}         \\ \toprule
        \multicolumn{2}{c}{Phoneme embedding dim.}                                                                                                                                      & 256                  & 384                        \\ \midrule
        \multirow{4}{*}{Phoneme-encoder}                                                               & \# of layers                                                                   & 4                    & 6                          \\
                                                                                                       & hidden dim.                                                                    & 256                  & 384                        \\
                                                                                                       & \# of kernels in Conv1D                                                        & 1024                 & 1536                       \\
                                                                                                       & kernel size in Conv1D                                                          & 9                    & 9                          \\ \midrule
        \multirow{4}{*}{Mel-decoder}                                                                   & \# of layers                                                                   & 4                    & 6                          \\
                                                                                                       & hidden dim.                                                                    & 256                  & 384                        \\
                                                                                                       & \# of kernels in Conv1D                                                        & 1024                 & 1536                       \\
                                                                                                       & kernel size in Conv1D                                                          & 9                    & 9                          \\ \midrule
        \multirow{2}{*}{\begin{tabular}[c]{@{}c@{}}Phoneme-encoder\\ to Mel-decoder\end{tabular}}    & \# of attention heads                                                          & 2                    & 2                          \\
                                                                                                       & dropout                                                                        & 0.1                  & 0.2                        \\ \midrule
        \multirow{4}{*}{Variance Predictor}                                                            & \# of Conv1D layers                                                          & 2                    & 2                          \\
                                                                                                       & \# of kernels in Conv1D                                                        & 256                  & 384                        \\
                                                                                                       & kernel size in Conv1D                                                          & 3                    & 3                          \\
                                                                                                       & dropout                                                                        & 0.5                  & 0.5                        \\ \midrule
        \multirow{5}{*}{\begin{tabular}[c]{@{}c@{}}Reference \\ Encoder\end{tabular}}                  & \# of Conv2D layers                                                            & -                    & 6                          \\
                                                                                                       & \# of kernels in Conv2D                                                        & -                    & (32, 32, 64, 64, 128, 128) \\
                                                                                                       & kernel size in Conv2D                                                          & -                    & (3, 3)                     \\
                                                                                                       & stride of Conv2D                                                               & -                    & (2, 2)                     \\
                                                                                                       & hidden dim. of GRU                                                             & -                    & 192                        \\ \midrule
        \multirow{4}{*}{\begin{tabular}[c]{@{}c@{}}Style Token \\ Layer\end{tabular}}                  & \# of tokens                                                                   & -                    & 10                         \\
                                                                                                       & token dimension                                                                & -                    & 48                         \\
                                                                                                       & \begin{tabular}[c]{@{}c@{}}hidden dim. of \\ multi-head-attention\end{tabular} & -                    & 384                        \\
                                                                                                       & \# of attention heads                                                          & -                    & 8                          \\ \midrule
        \multirow{3}{*}{\begin{tabular}[c]{@{}c@{}}Speaker/Emotion/ \\ Prosody Encoder\end{tabular}} & \# of kernels in Conv1D                                                        & -                    & 3                          \\
                                                                                                       & kernel size in Conv1D                                                          & -                    & 384                        \\
                                                                                                       & dropout                                                                        & -                    & 0.5                        \\ \toprule
        \multicolumn{2}{c}{\# of parameters}                                                                                                                                      & 27M                  & 94M                        \\ \bottomrule
                                                                                                       
        \end{tabular}
        \end{table}
    
\newpage

\section{Training procedure}

    We train UniTTS in three phases as follows:

\begin{enumerate}
    \item Train UniTTS activating the style encoder and deactivating the speaker, emotion, and prosody encoders.
    \item Train the speaker and emotion encoders.
        \begin{enumerate}
        \item Train the speaker and emotion embedding tables by distillation using the trained style encoder.
        \item Train the speaker, emotion encoders deactivating the style encoder and freezing the speaker and emotion embedding tables and other encoders.
        \end{enumerate}
    \item Train the phoneme-level prosody encoder freezing the other encoders.
\end{enumerate}

    The detailed procedure of each phase is presented in Algorithm \ref{algo:training_phase_1}-\ref{algo:training_phase_3}

    \begin{algorithm}[H]
        \SetAlgoLined

        \For {a predefined number of iterations} {
    
    
            \textbf{extract high-level embeddings from the input phonemes} \\
                \hspace{1em} $E_{phoneme} \gets PhonemeEncoder(y+PE)$
            
            \vspace{1em}
    
            \textbf{add style embedding}  $E_{phoneme} \gets E_{phoneme} + S(x_{se})$
    
            \vspace{1em}
    
            \textbf{predict phoneme durations}
                \hspace{1em} $\hat{D} \gets DurationPredictor(E_{phoneme})$ 
            
            \vspace{1em}

            \textbf{predict pitch} \\
                \hspace{1em} $\hat{Pitch} \gets PitchPredictor(E_{phoneme})$
            
            \vspace{0.3em}

            \textbf{add pitch embedding to the phoneme embedding} \\
                \hspace{1em} $E_{pitch} \gets PitchEncoder(Pitch_{GT} + shift_{pitch}, E_{phoneme})$ \\
                \hspace{1em} $E_{phoneme} \gets E_{phoneme} + E_{pitch}$  

            \vspace{1em}

            \textbf{predict energy} \\
                \hspace{1em} $\hat{Energy} \gets EnergyPredictor(E_{phoneme})$
            
            \vspace{0.3em}

            \textbf{add energy embedding to the phoneme embedding} \\
                \hspace{1em} $E_{energy} \gets EnergyEncoder(Energy_{GT} + shift_{energy}, E_{phoneme})$ \\
                \hspace{1em} $E_{phoneme} \gets E_{phoneme} + E_{energy}$ 
        
            \vspace{1em}        
            
            \textbf {align $E_{phoneme}$ by duplicating phoneme embeddings for the durations} \\
                \hspace{1em} $\tilde{E}_{phoneme} \gets LengthRegulator(E_{phoneme}, D_{GT})$
            
            \vspace{1em}        
            \textbf{synthesize spectrogram by decoder}  \\
                \hspace{1em} $\hat{x} \gets Decoder(\tilde{E}_{phoneme})$
            
            \vspace{1em}        
            \textbf{compute loss and backpropagate}  $L_{total} \gets L_{Mel} + L_{duration} + L_{pitch} + L_{energy}$ \\
                \hspace{1em} $L_{Mel} \gets MAE(x, \hat{x})$ \\
                \hspace{1em} $L_{duration} \gets MSE(D_{GT}, \hat{D})$ \\
                \hspace{1em} $L_{pitch} \gets MSE(Pitch_{GT}, \hat{Pitch})$ \\
                \hspace{1em} $L_{energy} \gets MSE(Energy_{GT}, \hat{Energy})$ \\
            
            \vspace{1em}        
        }
    
        \caption{Training phase \#1: Train UniTTS activating the style encoder and deactivating the speaker, emotion, and prosody encoders.}
        \label{algo:training_phase_1}
    \end{algorithm}
\newpage

    \begin{algorithm}[H]
        \SetAlgoLined
        // \textbf{Train speaker residual embedding table}
        
        \For {each speaker $s_{u} \in$ speaker set, $\{s_{u}\}$}{
            $\mu_{s_u} \gets \frac {1}{N_{s_u}} \sum_{s=s_u}S(x_{se})$
        }
        
        \vspace{1em}
        // \textbf{Train emotion residual embedding table}
        
        \For {each emotion-type $e_{v} \in$ emotion-type set, $\{e_{v}\}$}{
            $\mu_{e_v} \gets \frac {1}{N_{e_v}} \sum_{e=e_v} [S(x_{se}) - \mu_{s}]$
        }

        \vspace{1em}
        
        \For {a predefined number of iterations} {

            \textbf{extract high-level embeddings from the input phonemes} \\
                \hspace{1em} $E_{phoneme} \gets PhonemeEncoder(y+PE)$
            
            \vspace{1em}
                
            \textbf{add speaker embedding} \\
                \hspace{1em} $E_{spk} \gets sg(\mu_{s_u}) + SpeakerEncoder(sg(\mu_{s_u}), E_{phoneme})$ \\ 
                \hspace{1em} $E_{phoneme} \gets E_{phoneme} + E_{spk}$

            \vspace{1em}
            
            \textbf{add emotion embedding}  \\

                \hspace{1em} $E_{emo} \gets sg(\mu_{e_v}) + EmotionEncoder(sg(\mu_{e_v}), E_{phoneme})$ \\
                \hspace{1em} $E_{phoneme} \gets E_{phoneme} + E_{emo}$

            \vspace{1em}
            
            \textbf{predict phoneme durations}
                \hspace{1em} $\hat{D} \gets DurationPredictor(E_{phoneme})$ 
            
            \vspace{1em}

            \textbf{predict pitch} \\
                \hspace{1em} $\hat{Pitch} \gets PitchPredictor(E_{phoneme})$
            
            \vspace{0.3em}

            \textbf{add pitch embedding to the phoneme embedding} \\
                \hspace{1em} $E_{pitch} \gets PitchEncoder(Pitch_{GT} + shift_{pitch}, E_{phoneme})$ \\
                \hspace{1em} $E_{phoneme} \gets E_{phoneme} + E_{pitch}$  

            \vspace{1em}

            \textbf{predict energy} \\
                \hspace{1em} $\hat{Energy} \gets EnergyPredictor(E_{phoneme})$
            
            \vspace{0.3em}

            \textbf{add energy embedding to the phoneme embedding} \\
                \hspace{1em} $E_{energy} \gets EnergyEncoder(Energy_{GT} + shift_{energy}, E_{phoneme})$ \\
                \hspace{1em} $E_{phoneme} \gets E_{phoneme} + E_{energy}$ 
        
            \vspace{1em}        
            
            \textbf {align $E_{phoneme}$ by duplicating phoneme embeddings for the durations} \\
                \hspace{1em} $\tilde{E}_{phoneme} \gets LengthRegulator(E_{phoneme}, D_{GT})$
            
            \vspace{1em}        
            \textbf{synthesize spectrogram by decoder}  \\
                \hspace{1em} $\hat{x} \gets Decoder(\tilde{E}_{phoneme})$
            
            \vspace{1em}        
            \textbf{compute loss and backpropagate} $L_{total} \gets L_{Mel} + L_{duration} + L_{pitch} + L_{energy}$\\
                \hspace{1em} $L_{Mel} \gets MAE(x, \hat{x})$ \\
                \hspace{1em} $L_{duration} \gets MSE(D_{GT}, \hat{D})$ \\
                \hspace{1em} $L_{pitch} \gets MSE(Pitch_{GT}, \hat{Pitch})$ \\
                \hspace{1em} $L_{energy} \gets MSE(Energy_{GT}, \hat{Energy})$ \\
        }
        \caption{Training phase \#2: Train the speaker and emotion encoders.}
        \label{algo:training_phase_2}
    \end{algorithm}

    \newpage

    \begin{algorithm}[H]
        \SetAlgoLined
            
        \For {a predefined number of iterations} {

            \textbf{extract high-level embeddings from the input phonemes} \\
                \hspace{1em} $E_{phoneme} \gets PhonemeEncoder(y+PE)$
            
            \vspace{1em}
                
            \textbf{add speaker embedding} \\
                \hspace{1em} $E_{spk} \gets sg(\mu_{s_u}) + sg(SpeakerEncoder(sg(\mu_{s_u}), E_{phoneme}))$ \\ 
                \hspace{1em} $E_{phoneme} \gets E_{phoneme} + E_{spk}$

            \vspace{1em}
            
            \textbf{add emotion embedding}  \\

                \hspace{1em} $E_{emo} \gets sg(\mu_{e_v}) + sg(EmotionEncoder(sg(\mu_{e_v}), E_{phoneme}))$ \\
                \hspace{1em} $E_{phoneme} \gets E_{phoneme} + E_{emo}$

            \vspace{1em}
            
            \textbf{predict prosody embedding from phonemes (used in synthesis)} \\
                \hspace{1em} $\hat{E}_{prosody} \gets ProsodyPredictor(E_{phoneme})$ 
            
            \vspace{1em}
            
            \textbf{predict prosody embedding from Mel spectrogram }  \\
                \hspace{1em} $E_{prosody} \gets ProsodyEncoder(x_{mel-averaged-by-duration}, E_{phoneme})$ \\
            
            \vspace{1em}    
            
            \textbf{add prosody embedding}
                \hspace{1em} $E_{phoneme} \gets E_{phoneme} + E_{prosody}$
            
            \vspace{1em}
            
            \textbf{predict phoneme durations}
                \hspace{1em} $\hat{D} \gets DurationPredictor(E_{phoneme})$ 
            
            \vspace{1em}

            \textbf{predict pitch} \\
                \hspace{1em} $\hat{Pitch} \gets PitchPredictor(E_{phoneme})$
            
            \vspace{0.3em}

            \textbf{add pitch embedding to the phoneme embedding} \\
                \hspace{1em} $E_{pitch} \gets PitchEncoder(Pitch_{GT} + shift_{pitch}, E_{phoneme})$ \\
                \hspace{1em} $E_{phoneme} \gets E_{phoneme} + E_{pitch}$  

            \vspace{1em}

            \textbf{predict energy} \\
                \hspace{1em} $\hat{Energy} \gets EnergyPredictor(E_{phoneme})$
            
            \vspace{0.3em}

            \textbf{add energy embedding to the phoneme embedding} \\
                \hspace{1em} $E_{energy} \gets EnergyEncoder(Energy_{GT} + shift_{energy}, E_{phoneme})$ \\
                \hspace{1em} $E_{phoneme} \gets E_{phoneme} + E_{energy}$ 
        
            \vspace{1em}        
            
            \textbf {align $E_{phoneme}$ by duplicating phoneme embeddings for the durations} \\
                \hspace{1em} $\tilde{E}_{phoneme} \gets LengthRegulator(E_{phoneme}, D_{GT})$
            
            \vspace{1em}        
            \textbf{synthesize spectrogram by decoder}  \\
                \hspace{1em} $\hat{x} \gets Decoder(\tilde{E}_{phoneme})$
            
            \vspace{1em}        
            \textbf{compute loss and backpropagate} $L_{total} \gets L_{Mel} + L_{dur.} + L_{pitch} + L_{energy} + L_{pros.}$\\
                \hspace{1em} $L_{Mel} \gets MAE(x, \hat{x})$ \\
                \hspace{1em} $L_{dur.} \gets MSE(D_{GT}, \hat{D})$ \\
                \hspace{1em} $L_{pitch} \gets MSE(Pitch_{GT}, \hat{Pitch})$ \\
                \hspace{1em} $L_{energy} \gets MSE(Energy_{GT}, \hat{Energy})$ \\
                \hspace{1em} $L_{pros.} \gets MSE(E_{prosody}, \hat{E}_{prosody})$ \\        
        }
        
        \caption{Training phase \#3: Train the phoneme-level prosody encoder freezing the other encoders.}
        \label{algo:training_phase_3}
    \end{algorithm}

\newpage
\section{Speaker ID and emotion control}
        \begin{figure}[!htb]
            \centerline{\includegraphics[width=143mm]{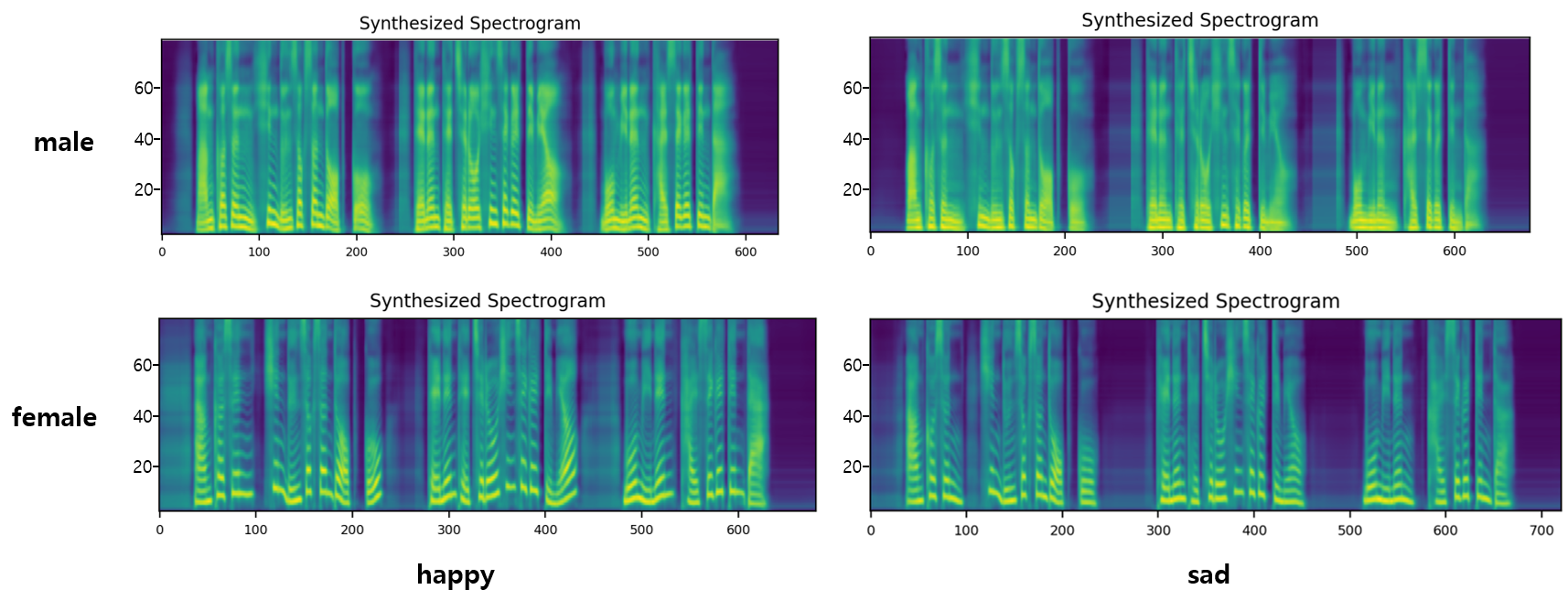}}
            \caption{Four spectrograms synthesized from the text “암컷은 흰 눈썹선도 없고, 배는 대체로 흰색을 띄며, 몸 위는 갈색을 띈다.” with different speaker and emotion IDs. Although the four spectrograms were synthesized from the same text, they show apparent differences, reflecting the different speaker IDs and emotion types. These figures exhibit that the proposed methods are effective in reflecting both speech ID and emotion.}
            \label{figure:spectrograms_from_the_same_text}
        \end{figure}

\section{Data augmentation}
        \begin{figure}[!htb]
            \centerline{\includegraphics[width=143mm]{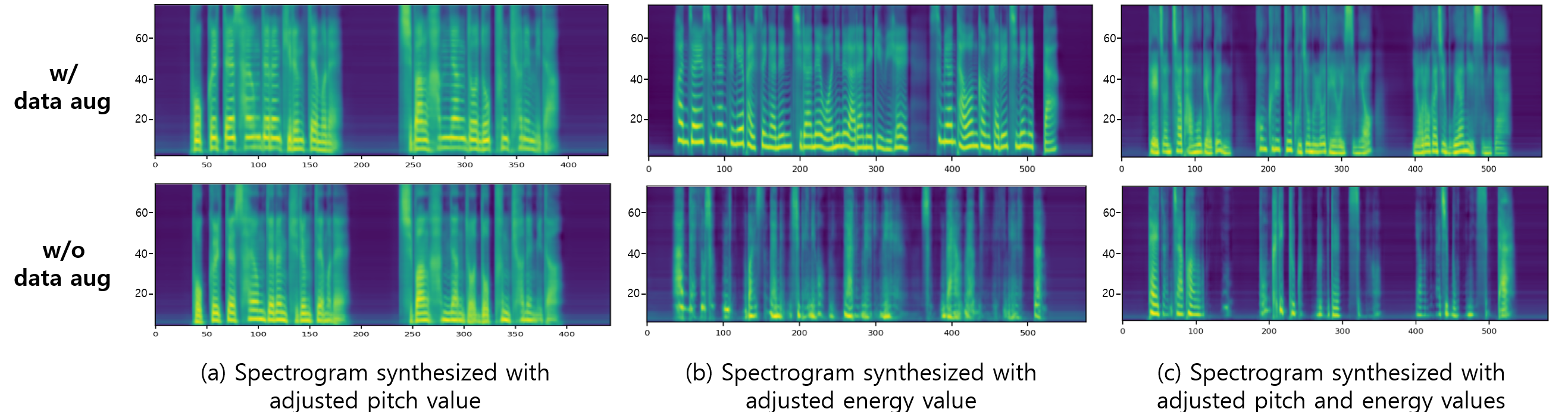}}
            \caption{The effect of the proposed data augmentation technique. The first and second rows display the spectrograms synthesized by the models trained with and without data augmentation, respectively. The spectrograms on the first row are clean while those on the second row were distorted. These figures exhibit that the proposed data augmentation technique is effective in improving fidelity and control over pitch and energy.}
            \label{figure:effect_data_aug}
        \end{figure}

\newpage

 \section{Style mixing}
        \begin{figure}[!htb]
            \centerline{\includegraphics[width=143mm]{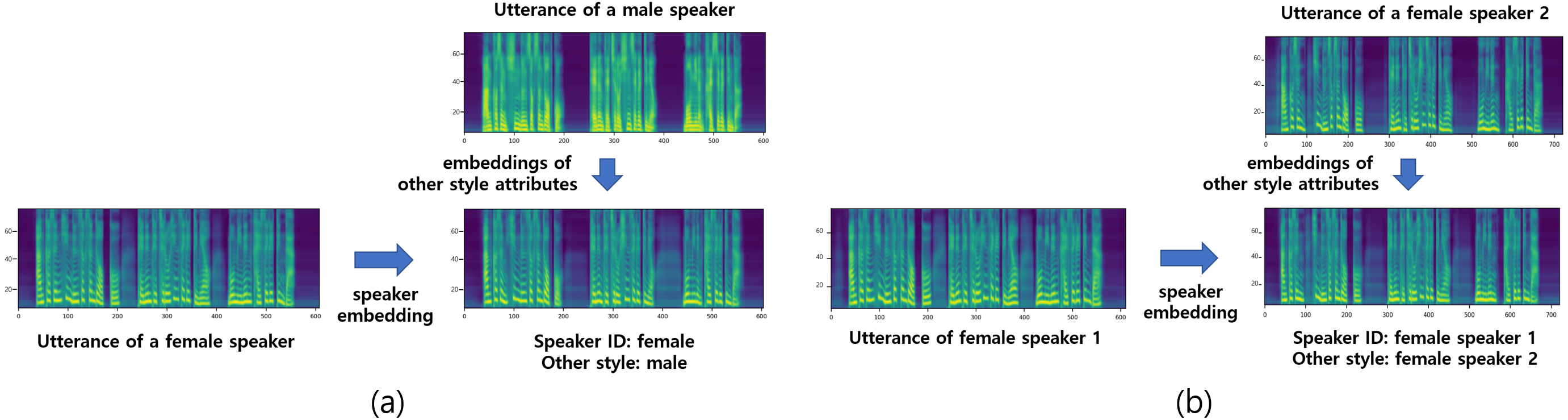}}
            \caption{The examples of style mixing results. We first synthesized two spectrograms using different speaker IDs, saving the speaker and other style attribute embeddings. Then, we synthesized a spectrogram mixing the saved speaker and style embeddings. In both (a) and (b), the lower right spectrograms were synthesized from the speaker embedding for the left spectrogram and the other style embeddings for the upper spectrogram. }
            \label{figure:style_mixing_examples}
        \end{figure}

\comment
{

\section{License of assets}
    \begin{table}[h]
        \centering
        \caption{License of assets used in this research}
        \medskip
        \begin{tabular}{@{}cc@{}}
        \toprule
        \textbf{Assets}                                       & \textbf{License}              \\    \midrule
        Sound eXchange (SoX) \cite{SoX}                       &  GNU Library or Lesser General Public v.2.0 \\
        Korean Single Speech (KSS) dataset \cite{KSS}         & CC-BY-NC-SA-4.0 International \\ 
        Korean Emotional Speech dataset \cite{KoreanEmotionalSpeech}    & CC-BY-NC-SA-4.0 International \\     
        EmotionTTS-Open-DB dataset \cite{EmotionTTS}          & CC-BY-NC-SA-4.0 International \\ 
        FastSpeech2 Opensource \cite{FastSpeech2_opensource}  & MIT License                   \\
        \bottomrule
        \end{tabular}
    \end{table}

\section{A note on the model name `UniTTS'}
    
    We are planning to change the name of our model from `UniTTS' to `UniTTS' because we found a previous work that uses the same name (\url{https://arxiv.org/pdf/2101.07597.pdf}). However, our work has no relation with the previous work, as the other `UniTTS' is a cross-lingual speech recognizer while our model is an expressive speech synthesizer. We apologize to the authors of the previous work for unintentional duplication of model names.

}   

\end{document}


\section*{Checklist}

The checklist follows the references.  Please
read the checklist guidelines carefully for information on how to answer these
questions.  For each question, change the default \answerTODO{} to \answerYes{},
\answerNo{}, or \answerNA{}.  You are strongly encouraged to include a {\bf
justification to your answer}, either by referencing the appropriate section of
your paper or providing a brief inline description.  For example:
\begin{itemize}
  \item Did you include the license to the code and datasets? \answerYes{See Section~\ref{gen_inst}.}
  \item Did you include the license to the code and datasets? \answerNo{The code and the data are proprietary.}
  \item Did you include the license to the code and datasets? \answerNA{}
\end{itemize}
Please do not modify the questions and only use the provided macros for your
answers.  Note that the Checklist section does not count towards the page
limit.  In your paper, please delete this instructions block and only keep the
Checklist section heading above along with the questions/answers below.

\begin{enumerate}

\item For all authors...
\begin{enumerate}
  \item Do the main claims made in the abstract and introduction accurately reflect the paper's contributions and scope?
    \answerYes{See Section 1. line 81.}
  \item Did you describe the limitations of your work?
    \answerYes{In Section 1 line 48, we mentioned that the proposed model is based on FastSpeech2.
              In Section 5.2.2 line 279, we mentioned that one of the proposed methods can decrease speaker similarity when the software toolkit changes timbre while adjusting the pitch of the voice. }
  \item Did you discuss any potential negative societal impacts of your work?
    \answerNo{ TTS technology can imitate other people's voices and can be used for inappropriate purposes. However, we were unable to suggest a non-obvious solution for the possibility of misuse. }
  \item Have you read the ethics review guidelines and ensured that your paper conforms to them?
    \answerYes{}
\end{enumerate}

\item If you are including theoretical results...
\begin{enumerate}
  \item Did you state the full set of assumptions of all theoretical results?
    \answerNA{ We propose new methods rather than theoretical results, and present experimental results exhibiting their effectiveness. }
  \item Did you include complete proofs of all theoretical results?
    \answerNA{ We propose new methods rather than theoretical results, and present experimental results exhibiting their effectiveness. }
\end{enumerate}

\item If you ran experiments...
\begin{enumerate}
  \item Did you include the code, data, and instructions needed to reproduce the main experimental results (either in the supplemental material or as a URL)?
    \answerNo{Instead, we present the details of the model structure, hyperparameter, and training procedure in appendix, and therefore, we believe many developer can reproduce our model using public dataset.}
  \item Did you specify all the training details (e.g., data splits, hyperparameters, how they were chosen)?
    \answerYes{See appendix.}
	\item Did you report error bars (e.g., with respect to the random seed after running experiments multiple times)?
    \answerNo{We do not present error rate. Instead, we present visualization results and MOS score.}
	\item Did you include the total amount of compute and the type of resources used (e.g., type of GPUs, internal cluster, or cloud provider)?
    \answerYes{See Subsection 5.1. "We ran the 250 experiments on a computer equipped with a Xeon E5-2630 v4 CPU and two NVIDIA GTX-1080Ti GPUs. The learning requires about one day when data augmentation was not applied, and about 4 days when applied."}
\end{enumerate}

\item If you are using existing assets (e.g., code, data, models) or curating/releasing new assets...
\begin{enumerate}
  \item If your work uses existing assets, did you cite the creators?
    \answerYes{See Subsection 5.1 and Reference 37-40}
  \item Did you mention the license of the assets?
    \answerYes{See Appendix.}
  \item Did you include any new assets either in the supplemental material or as a URL?
    \answerNo{}
  \item Did you discuss whether and how consent was obtained from people whose data you're using/curating?
    \answerNA{}
  \item Did you discuss whether the data you are using/curating contains personally identifiable information or offensive content?
    \answerNA{}
\end{enumerate}

\item If you used crowdsourcing or conducted research with human subjects...
\begin{enumerate}
  \item Did you include the full text of instructions given to participants and screenshots, if applicable?
    \answerNA{}
  \item Did you describe any potential participant risks, with links to Institutional Review Board (IRB) approvals, if applicable?
    \answerNA{}
  \item Did you include the estimated hourly wage paid to participants and the total amount spent on participant compensation?
    \answerNA{}
\end{enumerate}

\end{enumerate}


\newpage
\appendix
\section*{Appendix}

\section{The detailed structures of the encoders and predictors}
        \begin{figure}[h]
            \begin{subfigure}[b]{0.27\textwidth}
                 \centering
                 \includegraphics[width=\textwidth]{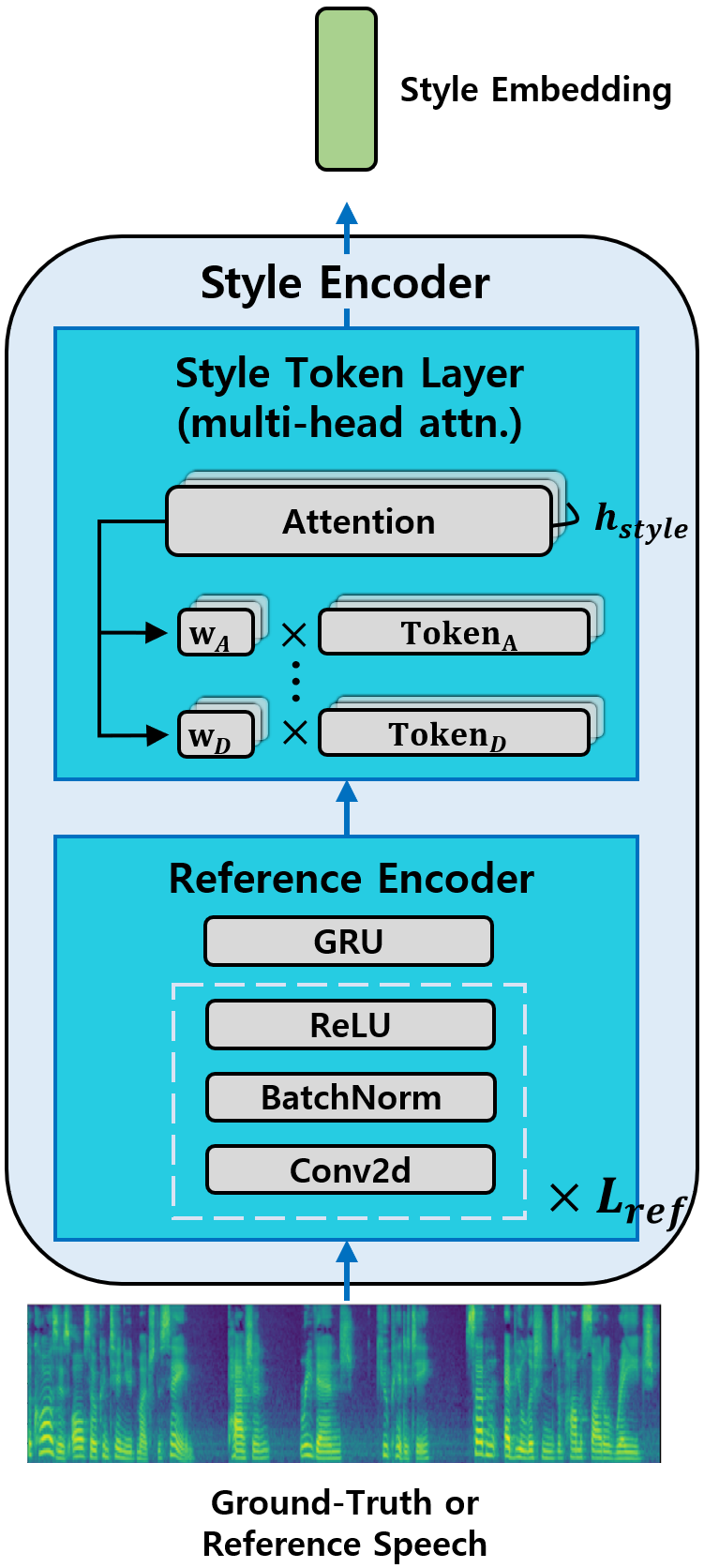}
                 \caption{Style encoder \newline }
            \end{subfigure}
            \hfill
            \begin{subfigure}[b]{0.2\textwidth}
                 \centering
                 \includegraphics[width=\textwidth]{assets/speaker_emotion_prosody_encoder_updated.png}
                 \caption{Speaker/emotion/ prosody encoder }
             \end{subfigure}
             \hfill
            \begin{subfigure}[b]{0.195\textwidth}    
                 \centering
                 \includegraphics[width=\textwidth]{assets/duration_predictor.png}
                 \caption{Duration predictor \newline  }
             \end{subfigure}
             \hfill
             \begin{subfigure}[b]{0.31\textwidth}
                 \centering
                 \includegraphics[width=\textwidth]{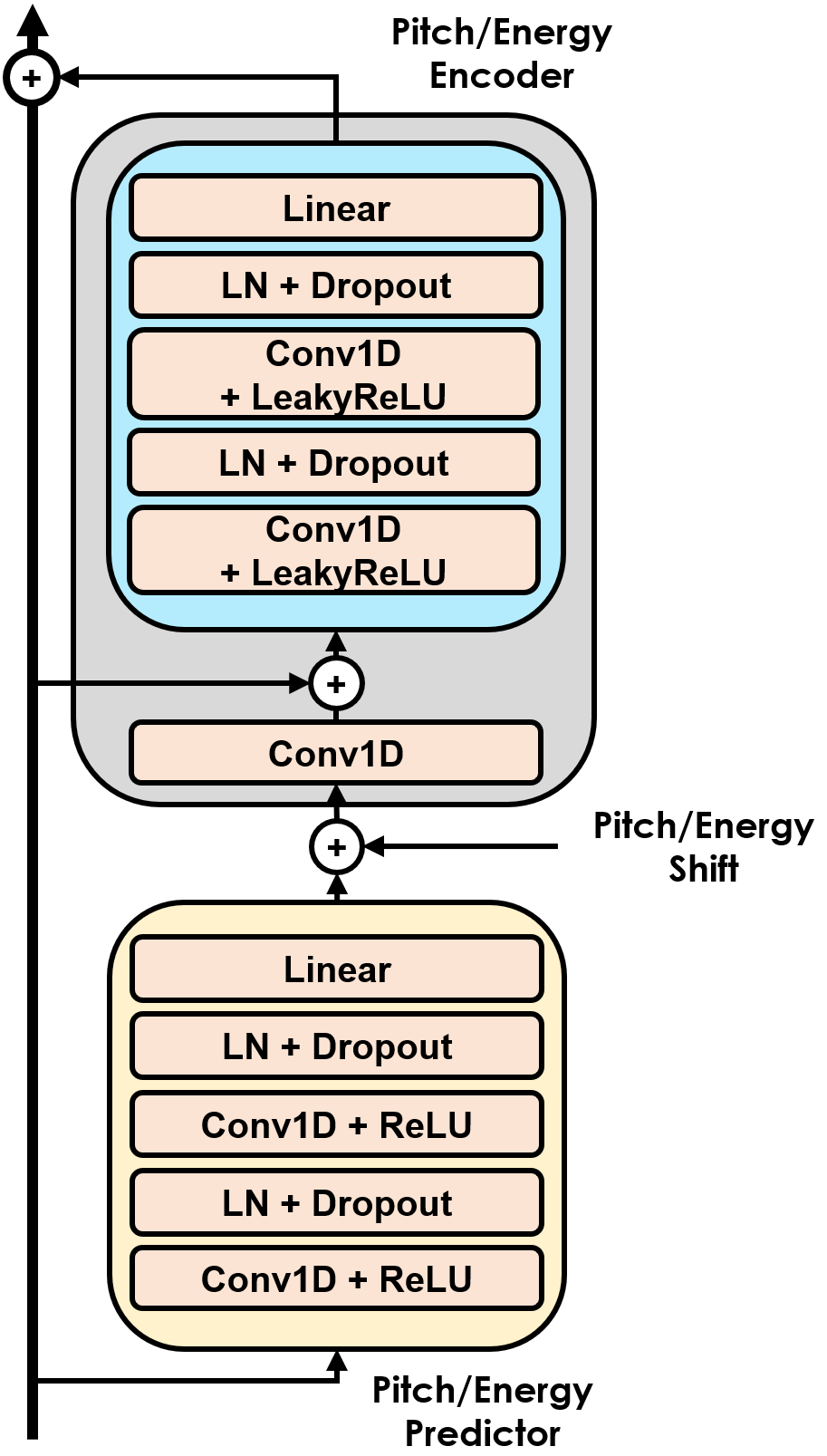}
                 \caption{Pitch/energy predictor \newline and encoder }
             \end{subfigure}             
     
            \caption{ The detailed structures of the predictors and encoders in Fig. 3. (a) The style encoder extracts a style embedding from a reference speech or ground-truth sample. It's structure is the same as that of [26], where $L_{ref}$ and $h_{style}$ are the number of Conv2d-BatchNorm-ReLU blocks and the number of attention heads, respectively. (b) The speaker/emotion encoder adapts the selected entry of the speaker/emotion embedding table by adding a residual vector. The prosody encoder outputs phone-level prosody embeddings. (c) The duration predictor predicts phoneme durations and pass them to the length regulator. (d) The pitch/energy predictor predicts the pitch/energy values of the phoneme embeddings. The pitch/energy encoder adds the pitch/energy embeddings to the phoneme embeddings. The proposed architecture allows adjusting pitch or energy by adding a pitch/energy shift to the predicted value. When we train the model with an augmented sample whose pitch or energy value was modified, we set the pitch or energy shift to the value used to augment the sample. This trick leads the model to learn to synthesize speech with a shifted pitch or energy value \cite{TransformingAutoEncoders}. }
            \label{fig:predictors_and_encoders}
        \end{figure}

    


\newpage

\section{Hyperparameters}

    \begin{table}[h]
        \centering
        \caption{The hyperparemeters of FastSpeech2 and UniSpeech}
        \medskip
        \begin{tabular}{@{}ccc@{}}
        \toprule
        \textbf{Hyperparameters}  & \textbf{FastSpeech2} & \textbf{UniSpeech} \\ \midrule
        
        Phoneme embedding dimension      & 256         & 384       \\
        
        Text-Encoder Layers              & 4           & 6         \\
        Text-Encoder Hidden              & 256         & 384       \\
        Text-Encoder Conv1D Kernel       & 9           & 9         \\ 
        Text-Encoder Conv1D Filter Size  & 1024        &1536       \\
        
        Mel-Decoder Layers               & 4           & 6         \\
        Mel-Decoder Hidden               & 256         & 384       \\
        Mel-Decoder Conv1D Kernel        & 9           & 9         \\
        Mel-Decoder Conv1D Filter Size   & 256         & 384       \\
        
        Text-Encoder/Mel-Decoder Attention Heads     & 2           & 2         \\
        Text-Encoder/Mel-Decoder Dropout & 0.1         & 0.2        \\
        
        Variance Predictor Conv1D Layers        & 2           & 2   \\
        Variance Predictor Conv1D Kernel & 3           & 3          \\
        Variance Predictor Conv1D Filter Size & 256       & 384     \\
        Variance Predictor Dropout      &   0.5         & 0.5       \\
        
        Reference Encoder Conv2D Layers & -             & 6        \\
        Reference Encoder Conv2D Kernel & -             & (3, 3)    \\
        Reference Encoder Conv2D Stride & -             & (2, 2)    \\
        Reference Encoder Conv2D Filter Sizes & -       & (32, 32, 64, 64, 128, 128) \\
        Reference Encoder GRU Hidden    & -             & 192 \\
        
        Style Encoder Token Number      & -             & 10        \\
        Style Encoder Token Dimension   &               & 48          \\
        Style Encoder Multihead Attention Hidden             & -             & 384         \\
        Style Encoder Multihead Attention Heads             & -             & 8         \\
        
        Speaker \& emotion \& prosody Encoder Conv1D Kernel  & -      & 3   \\
        Speaker \& emotion \& prosody Encoder Filter Size    & -      & 384  \\
        Speaker \& emotion \& prosody Encoder Dropout        & -      & 0.5 \\    
        
        Total Number of Parameters      & 27M           & 96M       \\
        \bottomrule
        \end{tabular}
        \label{Table:hyperparameters}
    \end{table}
    
\newpage

\section{Training procedure}

    We train UniSpeech in three phases as follows:

\begin{enumerate}
    \item Train UniSpeech activating the style encoder and deactivating the speaker, emotion, and prosody encoders.
    \item Train the speaker and emotion encoders.
        \begin{enumerate}
        \item Train the speaker and emotion embedding tables by distillation using the trained style encoder.
        \item Train the speaker, emotion encoders deactivating the style encoder and freezing the speaker and emotion embedding tables and other encoders.
        \end{enumerate}
    \item Train the phoneme-level prosody encoder freezing the other encoders.
\end{enumerate}

    The detailed procedure of each phase is presented in Algorithm \ref{algo:training_phase_1}-\ref{algo:training_phase_3}

    \begin{algorithm}[H]
        \SetAlgoLined

        \For {a predefined number of iterations} {
    
    
            \textbf{extract high-level embeddings from the input phonemes} \\
                \hspace{1em} $E_{phoneme} \gets PhonemeEncoder(y+PE)$
            
            \vspace{1em}
    
            \textbf{add style embedding}  $E_{phoneme} \gets E_{phoneme} + S(x_{se})$
    
            \vspace{1em}
    
            \textbf{predict phoneme durations}
                \hspace{1em} $\hat{D} \gets DurationPredictor(E_{phoneme})$ 
            
            \vspace{1em}

            \textbf{predict pitch} \\
                \hspace{1em} $\hat{Pitch} \gets PitchPredictor(E_{phoneme})$
            
            \vspace{0.3em}

            \textbf{add pitch embedding to the phoneme embedding} \\
                \hspace{1em} $E_{pitch} \gets PitchEncoder(Pitch_{GT} + shift_{pitch}, E_{phoneme})$ \\
                \hspace{1em} $E_{phoneme} \gets E_{phoneme} + E_{pitch}$  

            \vspace{1em}

            \textbf{predict energy} \\
                \hspace{1em} $\hat{Energy} \gets EnergyPredictor(E_{phoneme})$
            
            \vspace{0.3em}

            \textbf{add energy embedding to the phoneme embedding} \\
                \hspace{1em} $E_{energy} \gets EnergyEncoder(Energy_{GT} + shift_{energy}, E_{phoneme})$ \\
                \hspace{1em} $E_{phoneme} \gets E_{phoneme} + E_{energy}$ 
        
            \vspace{1em}        
            
            \textbf {align $E_{phoneme}$ by duplicating phoneme embeddings for the durations.} \\
                \hspace{1em} $\tilde{E}_{phoneme} \gets LengthRegulator(E_{phoneme}, D_{GT})$
            
            \vspace{1em}        
            \textbf{synthesize spectrogram by decoder}  \\
                \hspace{1em} $\hat{x} \gets Decoder(\tilde{E}_{phoneme})$
            
            \vspace{1em}        
            \textbf{compute loss and backpropagate}  $L_{total} \gets L_{Mel} + L_{duration} + L_{pitch} + L_{energy}$ \\
                \hspace{1em} $L_{Mel} \gets MAE(x, \hat{x})$ \\
                \hspace{1em} $L_{duration} \gets MSE(D_{GT}, \hat{D})$ \\
                \hspace{1em} $L_{pitch} \gets MSE(Pitch_{GT}, \hat{Pitch})$ \\
                \hspace{1em} $L_{energy} \gets MSE(Energy_{GT}, \hat{Energy})$ \\
            
            \vspace{1em}        
            
        }
    
        \caption{Training phase \#1: Train UniSpeech activating the style encoder and deactivating the speaker, emotion, and prosody encoders.}
        \label{algo:training_phase_1}
    \end{algorithm}
\newpage

    \begin{algorithm}[H]
        \SetAlgoLined
        // \textbf{Train speaker residual embedding table}
        
        \For {each speaker $s_{u} \in$ speaker set, $\{s_{u}\}$}{
            $\mu_{s_u} \gets \frac {1}{N_{s_u}} \sum_{s=s_u}S(x_{se})$
        }
        
        \vspace{1em}
        // \textbf{Train emotion residual embedding table}
        
        \For {each emotion-type $e_{v} \in$ emotion-type set, $\{e_{v}\}$}{
            $\mu_{e_v} \gets \frac {1}{N_{e_v}} \sum_{e=e_v} [S(x_{se}) - \mu_{s}]$
        }

        \vspace{1em}
        
        \For {a predefined number of iterations} {

            \textbf{extract high-level embeddings from the input phonemes} \\
                \hspace{1em} $E_{phoneme} \gets PhonemeEncoder(y+PE)$
            
            \vspace{1em}
                
            \textbf{add speaker embedding} \\
                \hspace{1em} $E_{spk} \gets sg(\mu_{s_u}) + SpeakerEncoder(sg(\mu_{s_u}), E_{phoneme})$ \\ 
                \hspace{1em} $E_{phoneme} \gets E_{phoneme} + E_{spk}$

            \vspace{1em}
            
            \textbf{add emotion embedding}  \\

                \hspace{1em} $E_{emo} \gets sg(\mu_{e_v}) + EmotionEncoder(sg(\mu_{e_v}), E_{phoneme})$ \\
                \hspace{1em} $E_{phoneme} \gets E_{phoneme} + E_{emo}$

            \vspace{1em}
            
            \textbf{predict phoneme durations}
                \hspace{1em} $\hat{D} \gets DurationPredictor(E_{phoneme})$ 
            
            \vspace{1em}

            \textbf{predict pitch} \\
                \hspace{1em} $\hat{Pitch} \gets PitchPredictor(E_{phoneme})$
            
            \vspace{0.3em}

            \textbf{add pitch embedding to the phoneme embedding} \\
                \hspace{1em} $E_{pitch} \gets PitchEncoder(Pitch_{GT} + shift_{pitch}, E_{phoneme})$ \\
                \hspace{1em} $E_{phoneme} \gets E_{phoneme} + E_{pitch}$  

            \vspace{1em}

            \textbf{predict energy} \\
                \hspace{1em} $\hat{Energy} \gets EnergyPredictor(E_{phoneme})$
            
            \vspace{0.3em}

            \textbf{add energy embedding to the phoneme embedding} \\
                \hspace{1em} $E_{energy} \gets EnergyEncoder(Energy_{GT} + shift_{energy}, E_{phoneme})$ \\
                \hspace{1em} $E_{phoneme} \gets E_{phoneme} + E_{energy}$ 
        
            \vspace{1em}        
            
            \textbf {align $E_{phoneme}$ by duplicating phoneme embeddings for the durations.} \\
                \hspace{1em} $\tilde{E}_{phoneme} \gets LengthRegulator(E_{phoneme}, D_{GT})$
            
            \vspace{1em}        
            \textbf{synthesize spectrogram by decoder}  \\
                \hspace{1em} $\hat{x} \gets Decoder(\tilde{E}_{phoneme})$
            
            \vspace{1em}        
            \textbf{compute loss and backpropagate} $L_{total} \gets L_{Mel} + L_{duration} + L_{pitch} + L_{energy}$\\
                \hspace{1em} $L_{Mel} \gets MAE(x, \hat{x})$ \\
                \hspace{1em} $L_{duration} \gets MSE(D_{GT}, \hat{D})$ \\
                \hspace{1em} $L_{pitch} \gets MSE(Pitch_{GT}, \hat{Pitch})$ \\
                \hspace{1em} $L_{energy} \gets MSE(Energy_{GT}, \hat{Energy})$ \\
        
        }
        
        \caption{Training phase \#2: Train the speaker and emotion encoders.}
        \label{algo:training_phase_2}
    \end{algorithm}

    \newpage

    \begin{algorithm}[H]
        \SetAlgoLined
            
        \For {a predefined number of iterations} {

            \textbf{extract high-level embeddings from the input phonemes} \\
                \hspace{1em} $E_{phoneme} \gets PhonemeEncoder(y+PE)$
            
            \vspace{1em}
                
            \textbf{add speaker embedding} \\
                \hspace{1em} $E_{spk} \gets sg(\mu_{s_u}) + sg(SpeakerEncoder(sg(\mu_{s_u}), E_{phoneme}))$ \\ 
                \hspace{1em} $E_{phoneme} \gets E_{phoneme} + E_{spk}$

            \vspace{1em}
            
            \textbf{add emotion embedding}  \\

                \hspace{1em} $E_{emo} \gets sg(\mu_{e_v}) + sg(EmotionEncoder(sg(\mu_{e_v}), E_{phoneme}))$ \\
                \hspace{1em} $E_{phoneme} \gets E_{phoneme} + E_{emo}$

            \vspace{1em}
            
            \textbf{predict prosody embedding from phonemes (used in synthesis)} \\
                \hspace{1em} $\hat{E}_{prosody} \gets ProsodyPredictor(E_{phoneme})$ 
            
            \vspace{1em}
            
            \textbf{predict prosody embedding from Mel spectrogram }  \\
                \hspace{1em} $E_{prosody} \gets ProsodyEncoder(x_{mel-averaged-by-duration}, E_{phoneme})$ \\
            
            \vspace{1em}    
            
            \textbf{add prosody embedding}
                \hspace{1em} $E_{phoneme} \gets E_{phoneme} + E_{prosody}$
            
            \vspace{1em}
            
            \textbf{predict phoneme durations}
                \hspace{1em} $\hat{D} \gets DurationPredictor(E_{phoneme})$ 
            
            \vspace{1em}

            \textbf{predict pitch} \\
                \hspace{1em} $\hat{Pitch} \gets PitchPredictor(E_{phoneme})$
            
            \vspace{0.3em}

            \textbf{add pitch embedding to the phoneme embedding} \\
                \hspace{1em} $E_{pitch} \gets PitchEncoder(Pitch_{GT} + shift_{pitch}, E_{phoneme})$ \\
                \hspace{1em} $E_{phoneme} \gets E_{phoneme} + E_{pitch}$  

            \vspace{1em}

            \textbf{predict energy} \\
                \hspace{1em} $\hat{Energy} \gets EnergyPredictor(E_{phoneme})$
            
            \vspace{0.3em}

            \textbf{add energy embedding to the phoneme embedding} \\
                \hspace{1em} $E_{energy} \gets EnergyEncoder(Energy_{GT} + shift_{energy}, E_{phoneme})$ \\
                \hspace{1em} $E_{phoneme} \gets E_{phoneme} + E_{energy}$ 
        
            \vspace{1em}        
            
            \textbf {align $E_{phoneme}$ by duplicating phoneme embeddings for the durations.} \\
                \hspace{1em} $\tilde{E}_{phoneme} \gets LengthRegulator(E_{phoneme}, D_{GT})$
            
            \vspace{1em}        
            \textbf{synthesize spectrogram by decoder}  \\
                \hspace{1em} $\hat{x} \gets Decoder(\tilde{E}_{phoneme})$
            
            \vspace{1em}        
            \textbf{compute loss and backpropagate} $L_{total} \gets L_{Mel} + L_{dur.} + L_{pitch} + L_{energy} + L_{pros.}$\\
                \hspace{1em} $L_{Mel} \gets MAE(x, \hat{x})$ \\
                \hspace{1em} $L_{dur.} \gets MSE(D_{GT}, \hat{D})$ \\
                \hspace{1em} $L_{pitch} \gets MSE(Pitch_{GT}, \hat{Pitch})$ \\
                \hspace{1em} $L_{energy} \gets MSE(Energy_{GT}, \hat{Energy})$ \\
                \hspace{1em} $L_{pros.} \gets MSE(E_{prosody}, \hat{E}_{prosody})$ \\        
        }
        
        \caption{Training phase \#3: Train the phoneme-level prosody encoder freezing the other encoders.}
        \label{algo:training_phase_3}
    \end{algorithm}

\newpage
\section{Speaker ID and emotion control}
        \begin{figure}[!htb]
            \centerline{\includegraphics[width=143mm]{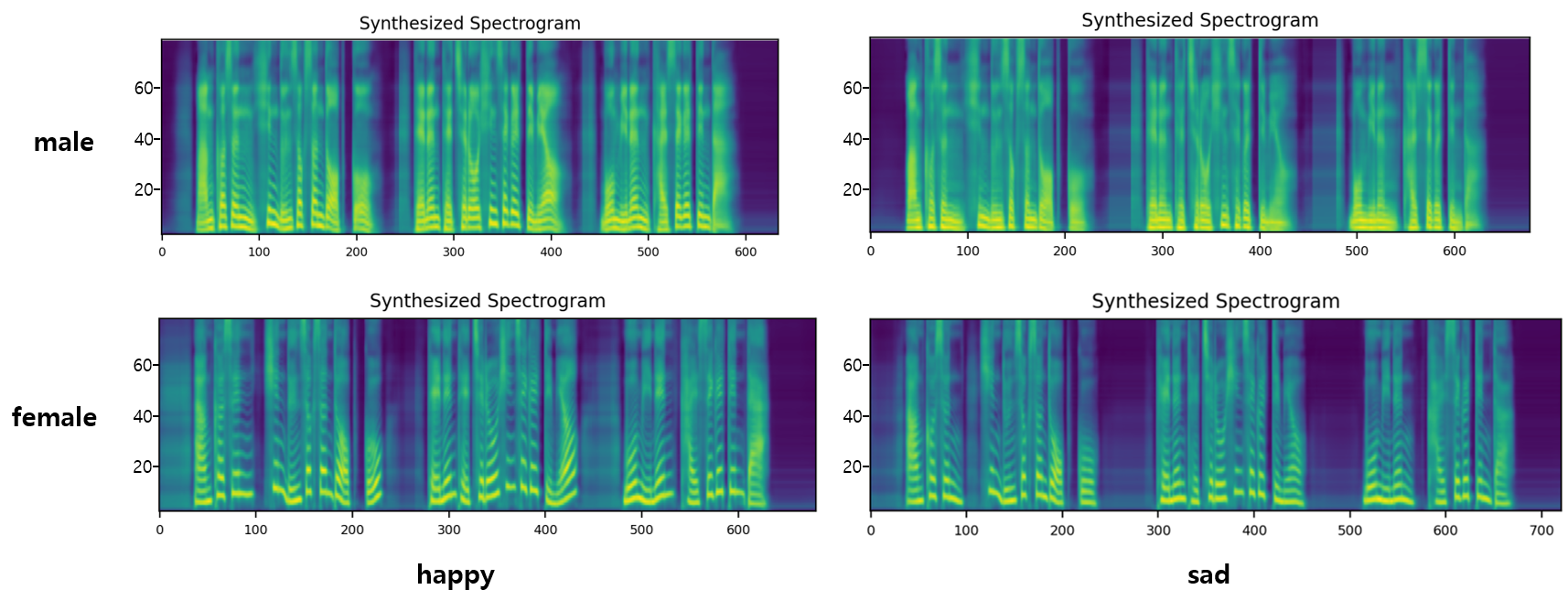}}
            \caption{Four spectrograms synthesized from the text “암컷은 흰 눈썹선도 없고, 배는 대체로 흰색을 띄며, 몸 위는 갈색을 띈다.” with different speaker and emotion IDs. Although the four spectrograms were synthesized from the same text, they show apparent differences, reflecting the different speaker IDs and emotion types. These figures exhibit that the proposed methods are effective in reflecting both speech ID and emotion.}
            \label{figure:spectrograms_from_the_same_text}
        \end{figure}

\section{Data augmentation}
        \begin{figure}[!htb]
            \centerline{\includegraphics[width=143mm]{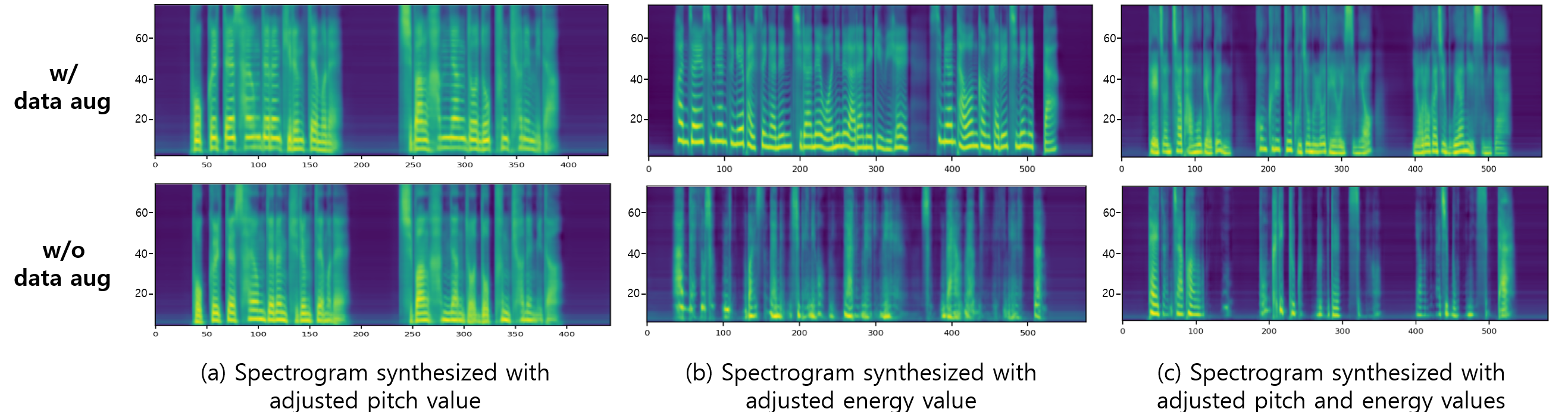}}
            \caption{The effect of the proposed data augmentation technique. The first and second rows display the spectrograms synthesized by the models trained with and without data augmentation, respectively. The spectrograms on the first row are clean while those on the second row were distorted. These figures exhibit that the proposed data augmentation technique is effective in improving fidelity and control over pitch and energy.}
            \label{figure:effect_data_aug}
        \end{figure}

\newpage

 \section{Style mixing}
        \begin{figure}[!htb]
            \centerline{\includegraphics[width=143mm]{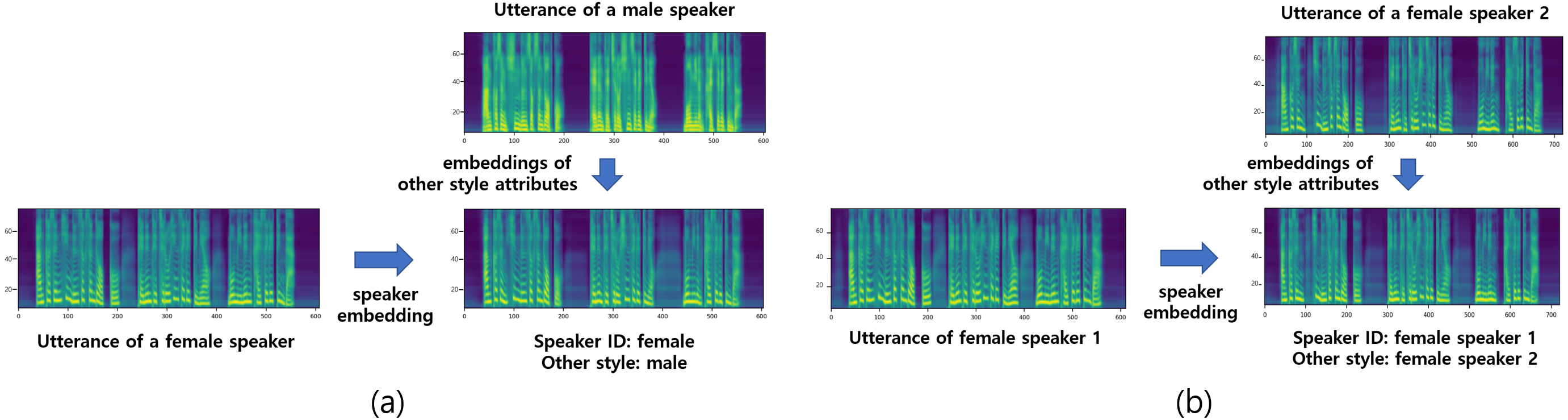}}
            \caption{The examples of style mixing results. We first synthesized two spectrograms using different speaker IDs, saving the speaker and other style attribute embeddings. Then, we synthesized a spectrogram mixing the saved speaker and style embeddings. In both (a) and (b), the lower right spectrograms were synthesized from the speaker embedding for the left spectrogram and the other style embeddings for the upper spectrogram. }
            \label{figure:style_mixing_examples}
        \end{figure}

\section{License of assets}
    \begin{table}[h]
        \centering
        \caption{License of assets used in this research}
        \medskip
        \begin{tabular}{@{}cc@{}}
        \toprule
        \textbf{Assets}                                       & \textbf{License}              \\    \midrule
        Sound eXchange (SoX) [36]                        &  GNU Library or Lesser General Public v.2.0 \\
        Korean Single Speech (KSS) dataset [37]          & CC-BY-NC-SA-4.0 International \\ 
        Korean Emotional Speech dataset [38]    & CC-BY-NC-SA-4.0 International \\     
        EmotionTTS-Open-DB dataset [39]         & CC-BY-NC-SA-4.0 International \\ 
        FastSpeech2 Opensource [40]  & MIT License                   \\
        \bottomrule
        \end{tabular}
    \end{table}

\section{A note on the model name `UniSpeech'}
    
    We are planning to change the name of our model from `UniSpeech' to `UniTTS' because we found a previous work that uses the same name (\url{https://arxiv.org/pdf/2101.07597.pdf}). However, our work has no relation with the previous work, as the other `UniSpeech' is a cross-lingual speech recognizer while our model is an expressive speech synthesizer. We apologize to the authors of the previous work for unintentional duplication of model names.